\documentclass[12pt,preprint]{aastex}

\usepackage{graphicx}
\usepackage{enumerate}
\usepackage{amsmath}

\def\orch{\textit{Orchestra}}
\def\orchestra{\orch}
\def\nbody{$n$-body}
\def\ms{m/s}
\def\ergg{erg/g}

\def\newhorizons{\textit{New Horizons}}
\def\cassini{\textit{Cassini}}

\def\mumolec{\mu_{\rm mol}}
\def\mumH{\mumolec m_{\rm H}}
\def\rmolec{r_{\rm mol}}
\def\mfp{\lambda}
\def\vtherm{v_{\rm therm}}
\def\gammaadiabat{\gamma}
\def\rhogas{\rho_{\rm gas}}
\def\Mgas{M_{\rm gas}}
\def\Msats{M_{\rm snkh}}
\def\cs{c_{\rm s}}
\def\Nfifty{N_{50}}
\def\fkb{x_{\rm KB}}

\def\adrag{a_{\rm drag}}
\def\vej{v_{\rm ej}}
\def\vmin{v_{\rm min}}
\def\vesc{v_{\rm esc}}
\def\vkep{v_{_{\rm K}}}

\def\tdamp{t_{\rm damp}}
\def\tdecay{t_{\rm decay}}
\def\tboost{t_{\rm boost}}

\def\tcol{t_{\rm col}}
\def\tcoloh{t_0}
\def\tstop{t_\text{stop}}

\def\MP{M_{\rm P}}
\def\RP{\ifmmode{R_{\rm P}}\else{$R_{\rm P}$}\fi}
\def\MC{M_{\rm C}}
\def\RC{R_{\rm C}}
\def\rhoc{\rho_{\rm C}}
\def\abin{{a_{\rm bin}}}
\def\Dasatzone{\Delta a}

\def\rp{r}
\def\vp{v}
\def\vprel{v_{\rm rel}}
\def\rhop{\rho}

\def\rP{\tilde{r}}
\def\mP{\tilde{m}}
\def\aP{\tilde{a}}
\def\eP{\tilde{e}}

\def\vPrel{\tilde{v}_{\rm rel}}
\def\rhoP{\tilde{\rho}}

\def\qdstar{Q_D^\ast}
\def\rdstar{r_{\!D}^\ast}

\def\ieff{\imath_{\rm eff}}
\def\eeff{e_{\rm eff}}

\def\vrel{v_{\rm rel}}

\def\fd{f}
\def\gd{g}
\def\gb{g_b}

\def\Mi{M_\text{im}}
\def\ri{r_\text{im}}
\def\vi{v_\text{im}}
\def\thetai{\theta_\text{im}}
\def\rhoi{\rho_\text{im}}

\def\dsp{\displaystyle}

\def\mytxt#1{\mbox{{\small #1}}}
\def\myvar#1{\mbox{{${#1}$}}}

\def\Mgasnineteen{\left[\frac{\Mgas}{\mytxt{$10^{19}$ g}}\right]}
\def\Tforty{\left[\frac{T}{\mytxt{40 K}}\right]}
\def\mumHwater{\left[\frac{\myvar{\mumH}}{\mytxt{18}}\right]}
\def\rmolecNtwo{\left[\frac{\rmolec}{\mytxt{0.3 nm}}\right]}
\def\Sonecgs{\dsp\left[\frac{\myvar{\Sigma}}{\mytxt{1 g/cm$^2$}}\right]}
\def\Dathirtythree{\dsp\left[\frac{\myvar{\Dasatzone}}{\mytxt{$33 \RP$}}\right]}

\def\qbtwoesix{\dsp\left[\frac{\myvar{Q_b}}{\mytxt{$2\times 10^6$ erg/g$\cdot$cm$^{0.4}$}}\right]}
\def\afiftyRp{\dsp\left[\frac{\myvar{a}}{\mytxt{$50 \RP$}}\right]}
\def\itwentyfive{\dsp\left[\frac{\myvar{\ieff}}{\mytxt{$25^\circ$}}\right]}
\def\ifive{\dsp\left[\frac{\myvar{\ieff}}{\mytxt{$5^\circ$}}\right]}
\def\eohseven{\dsp\left[\frac{\myvar{\eeff}}{\mytxt{0.7}}\right]}
\def\rptenm{\dsp\left[\frac{\myvar{\rp}}{\mytxt{10 m}}\right]}
\def\rponecm{\dsp\left[\frac{\myvar{\rp}}{\mytxt{1 cm}}\right]}
\def\rponemm{\dsp\left[\frac{\myvar{\rp}}{\mytxt{1 mm}}\right]}
\def\rhoponecgs{\dsp\left[\frac{\myvar{\rhop}}{\mytxt{1 g/cm$^3$}}\right]}

\def\aPthirtyRp{\dsp\left[\frac{\myvar{\aP}}{\mytxt{$30 \RP$}}\right]}
\def\opefac{\frac{1+\eP/2}{1.25}}
\def\aPonehundredRp{\dsp\left[\frac{\myvar{\aP}}{\mytxt{$100 \RP$}}\right]}
\def\rhoPonecgs{\dsp\left[\frac{\myvar{\rhoP}}{\mytxt{1 g/cm$^3$}}\right]}
\def\rPtenm{\dsp\left[\frac{\myvar{\rP}}{\mytxt{10 m}}\right]}

\def\MvtenMsats{\dsp\left[\frac{\myvar{M(v)}}{\myvar{10 \Msats}}\right]}

\def\vitwokms{\dsp\left[\frac{\myvar{\vi}}{\mytxt{2 km/s}}\right]}

\begin{document}

\title{A Pluto-Charon Concerto: An Impact on Charon as the
  Origin of the Small Satellites}

\author{Benjamin C. Bromley}
\affil{Department of Physics \& Astronomy, University of Utah, 
\\ 115 S 1400 E, Rm 201, Salt Lake City, UT 84112}
\email{bromley@physics.utah.edu}

\author{Scott J. Kenyon}
\affil{Smithsonian Astrophysical Observatory,
\\ 60 Garden St., Cambridge, MA 02138}
\email{skenyon@cfa.harvard.edu}

\begin{abstract}
{We consider a scenario where the small satellites of Pluto and
  Charon grew within a disk of debris from an impact between Charon
  and a trans-Neptunian Object (TNO). After Charon's orbital motion
  boosts the debris into a disk-like structure, rapid orbital damping
  of meter-size or smaller objects is essential to prevent the
  subsequent re-accretion or dynamical ejection by the binary.  From
  analytical estimates and simulations of disk evolution, we estimate
  an impactor radius of 30--100~km; smaller (larger) radii apply to an
  oblique (direct) impact.  Although collisions between large TNOs and
  Charon are unlikely today, they were relatively common within the
  first 0.1--1~Gyr of the solar system. Compared to models where the
  small satellites agglomerate in the debris left over by the giant
  impact that produced the Pluto-Charon binary planet, satellite
  formation from a later impact on Charon avoids the destabilizing
  resonances that sweep past the satellites during the early orbital
  expansion of the binary.  }

\end{abstract}


\section{Introduction}

The spectacular \newhorizons\ flyby of Pluto and Charon has deepened
the mystery surrounding the binary planet's delicate system of
satellites.  With orbital periods close to resonances at 3:1 (Styx),
4:1 (Nix), 5:1 (Kerberos), and 6:1 (Hydra) times the 6.4-d period of
the central binary, the satellites are as tightly packed as possible
\citep[e.g.,][]{weaver2006, buie2006, tholen2008, youdin2012,
  showalter2015, brozovic2015, kb2019a, kb2019b}. Despite their
intriguing orbits, the satellites seem like an afterthought, together
accounting for less than 0.001\% of Pluto's mass. How these moons
formed remains uncertain.

In the most popular formation model, the small moons arose from debris
after the grazing collision that led to the formation of the
Pluto-Charon binary \citep{canup2005, ward2006, asphaug2006,
  canup2011, desch2015, mckinnon2017}.  While this event ejects enough
material to make the small moons, the central binary probably acquires
an eccentric orbit with a period of 1--2~days soon after the
collision.  Driven by tidal interactions, the orbits of Pluto and
Charon circularize and drift apart with time.  As the binary expands,
the locations of $n:1$ resonances with the binary orbit sweep outward
and destabilize circumbinary material \citep[e.g.,][]{lith2008a,
  cheng2014b, smullen2017, woo2018}. Although collisional damping and
gravity-driven viscosity might stabilize small particles within
resonances \citep{bk2015pc}, it is not clear whether these particles
can collide and merge into larger satellites that remain on stable
orbits as the central binary evolves.

In another scenario, Charon grows within a massive debris swarm
produced from a collision with Pluto that largely destroyed the
impactor \citep{canup2001, canup2005, kb2019c}. As with a grazing
collision, the debris from the impact provides a reservoir for the
small satellites.  Again, uncertainties in the subsequent evolution of
the system allow only speculation that the four moons could survive
the tidal expansion of the binary.

Here, we explore the idea that the satellites grew out of debris from
a giant impact well after the binary planet settled into its present
configuration.  In this picture, ejecta from a collision between a TNO
and Charon forms a flattened, highly eccentric, circumbinary swarm of
solids.  The challenge is to find a mechanism that rapidly damps the
orbits of debris particles before they are accreted or ejected by the
binary.  After reviewing the Pluto-Charon system (\S2), we provide
analytical estimates to compare different mechanisms for dynamically
cooling debris orbits (\S3), bolstered by simulations of disk
evolution with the hybrid \nbody-coagulation code, \orch\ (\S4).
Then, in \S5, we use our results to specify broad requirements for the
successful production of the small moons by an impact with Charon. Our
conclusions are in \S6.

\section{Preliminaries}

\begin{figure}[t!]
\centerline{\includegraphics[width=5in]{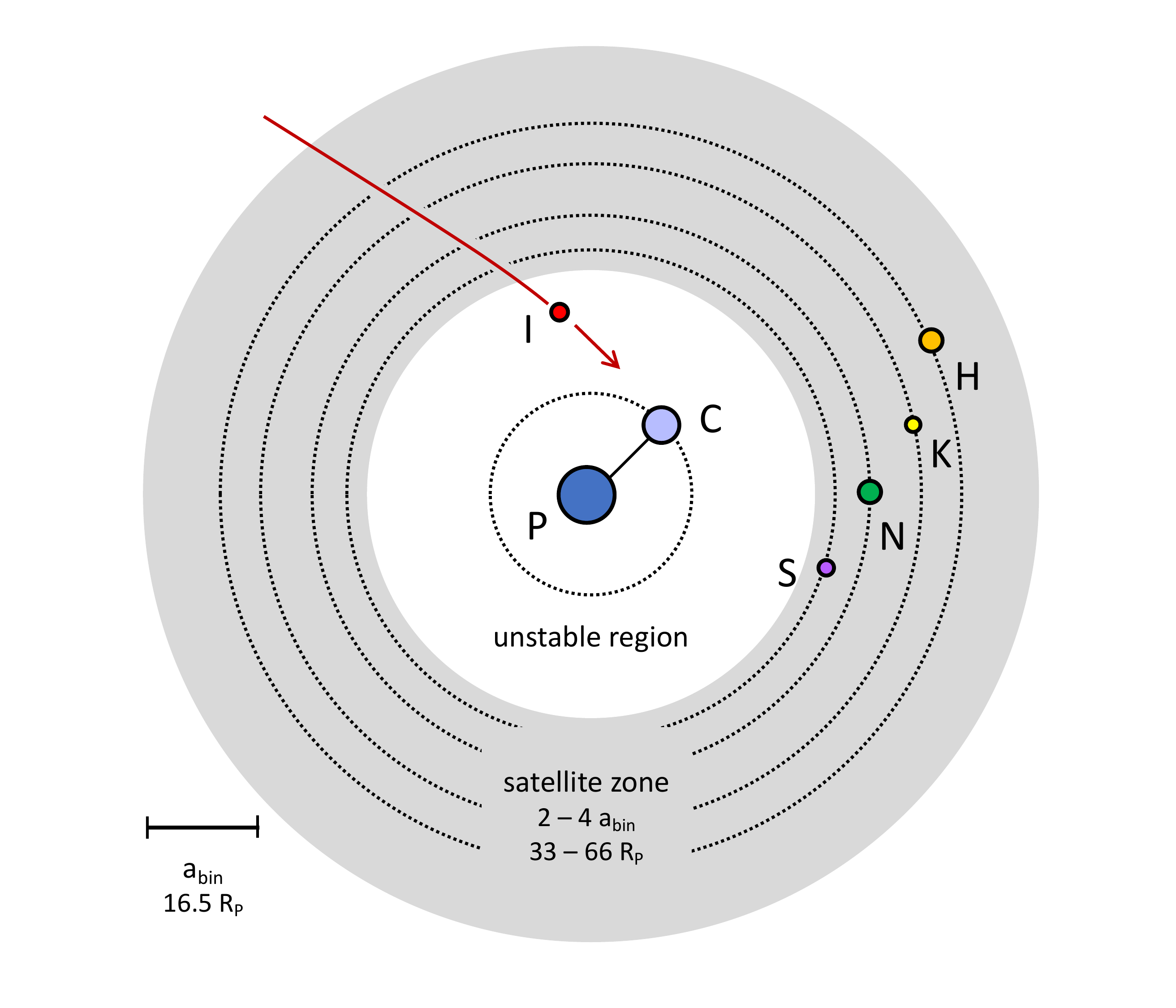}}
\caption{\label{fig:pc} A schematic of the Pluto-Charon satellite
  system with a hypothetical trans-Neptunian impactor. The orbits
  are to scale, while the sizes of the symbols are suggestive of
  physical sizes only; ``PC'' are the binary, ``SNKH'' are the small
  satellites, and ``I'' designates an impactor that may scatter debris
  into the satellite zone. Debris particles that manage to settle on
  orbits in the unstable region are ejected or accreted by the
  binary.}
\end{figure}

Today, the small satellites of Pluto and Charon lie on nearly circular
orbits in the plane of the binary at distances between 35$\RP$ and
55$\RP$ from the barycenter, where $\RP = 1188$~km is Pluto's radius
\citep{stern2015, stern2018}.  We define a ``satellite zone'' that
encompasses these orbits, an annular region extending from 33$\RP$ to
66$\RP$ around Pluto-Charon's center of mass, a range that is
equivalent to 2--4 times the binary separation, $\abin \approx
16.5~\RP\ \approx 19,600$~km.  The midplane of this zone coincides
with the orbital plane of the central binary and the satellite system.
Interior to this zone, prograde, coplanar orbits around the binary are
unstable \citep[e.g.,][]{dvorak1989, holman1999, doolin2011,
  youdin2012, kb2019a, gaslacgallardo2019}.  Styx, near the inner edge
of the satellite zone, seems unnervingly close to this unstable
region. Figure~\ref{fig:pc} and Table~\ref{tab:pcsystem} together
summarize the present-day configuration of the satellite system.

\begin{table}[t!]
\caption{\label{tab:pcsystem} Nominal properties of the Pluto-Charon
  system\tablenotemark{a}}
\begin{center}
\begin{tabular}{lccccc}
\hline
\hline 
Name & Mass (kg) & Radius (km) & $a/\abin$  & $a/\RP$ & Period (d)
\\
\hline
Pluto & $1.303\times 10^{25}$ & 1188  &---&---& 6.39
\\
Charon & $1.587\times 10^{24}$ & 606 & 1 & 16.49 & 6.39
\\
Styx & $\lesssim 5\times 10^{18}$ & 5.2 & 2.178 & 35.94 & 20.2
\\
Nix  & $45\times 10^{18}$ & 19.3 & 2.485 & 41.02 & 24.9
\\
Kerberos & $\lesssim 16\times 10^{18}$ & 6 & 2.949 & 48.681  & 32.2
\\
Hydra & $48\times 10^{18}$ & 20.9 & 3.303 & 54.54 & 38.2
\\
Satellite zone & $1.2\times 10^{20}$ & ($> 30$) & 1--2 & 33--66 & 17.8--51.0
\\
\hline
\end{tabular}
\\
\parbox{4in}{
{\footnotesize $^a$See \citet{brozovic2015}, \citet{stern2015},
    \citet{weaver2016}, \citet{nimmo2017}, and \citet{mckinnon2017},
     and \citet{kb2019b}.}}
\end{center}
\end{table}
  
The mass in the satellite zone is at least the sum total of the masses
of the four satellites, $\Msats$. On the basis of \textit{Hubble Space
  Telescope} and \newhorizons\ observations \citep{weaver2006,
  showalter2011, showalter2012, brozovic2015, weaver2016}, along with
dynamical studies \citep{youdin2012, showalter2015, kb2019b}, we adopt
$\Msats = 1.2\times 10^{20}$~g, the mass of a modest-size (30~km) icy
body (Table~\ref{tab:pcsystem}).  This estimate provides a guideline
for establishing the mass that must be delivered to the satellite zone
by a major collision with the binary.

Our goal here is to assess whether an impact between a TNO and the
Pluto-Charon binary leads to the formation of the four satellites.  
We focus on Charon as the target since its orbital motion helps boost
impact debris into a prograde circumbinary disk (\S\ref{sec:formdisk}). 
%
%
We assess how efficiently mass from an impact is delivered to the satellite
zone (\S3) and how it settles there in a dynamically cool disk (\S4).
These calculations informs the mass of the impactor and 
the impact geometry (\S\ref{sec:discuss}). If the impact is a direct hit, 
as in a cratering event, then the estimated projectile mass is roughly 
ten times $\Msats$, corresponding to a $\sim$100~km TNO. A smaller 
projectile may deliver enough material directly to the satellite if its 
impact is a well-aimed, surface-skimming event \citep{lein2012}, beaming
debris into a prograde orbit.  In any case, material launched into the
satellite zone must somehow settle there. We place constraints on this
process next.

\section{Formation of a circumbinary disk}\label{sec:formdisk}

An impact between a TNO and either Pluto or Charon produces copious
amounts of debris. In this section we explore how this material is
delivered to the satellite zone, and how it might settle onto stable
orbits in the plane of the binary. The main results include estimates
of debris particle sizes and total debris mass required to place
enough material in the satellite zone for building Nix and its
siblings.  We begin with an argument that binary dynamics make it
easier to deliver debris to the satellite zone, and that Charon is the
better target for the TNO impact.

\subsection{Impact events on a binary: Why Charon} 

An impact event on the surface of a single planet---as in Pluto before
it was joined by Charon---can lead to a range of outcomes, depending
on the impact parameter and impactor size, among other factors. In a
simple scenario, the impact obliterates the projectile, kicking up
many small debris particles that do not interact with each other.
Particles moving faster than the planet's escape speed are lost;
slower-moving objects fall back onto the planet's surface.  Through
ejection and accretion, all of the debris is gone in a dynamical time.

A binary partner like Charon changes this picture. Because both
planets are in motion relative to the center of mass, they are not
always easy targets for debris that falls back toward them.  Instead,
many debris particles make multiple close passes by the binary before
being accreted or ejected. Debris particles also have significant
angular momentum in the center-of-mass frame if they are launched from
the surface of the secondary and are boosted by its orbital motion
about the primary.  The result is a disk-like swarm of debris, roughly
aligned with the binary's orbital plane, that
survives much longer than the dynamical time.

{The presence of a binary partner also determines how the spray
  of collision ejecta eventually settles into a disk. As seen in the early work
  of \citet{brah1975, brah1976}, debris particles around a point mass
  will dynamical cool through collisional processes into a common
  midplane established by their total angular momentum. The presence
  of a massive binary partner like Charon produces tidal torques that
  coerce debris into the plane of the binary
  \citep[e.g.,][]{larwood1997, foucart2013}, similar to the effect of
  Saturn's axisymmetric potential on its thin, coplanar rings
  \citep{gold1982}. When impact debris is launched from the secondary,
  the net angular momentum of the spray of ejecta is already roughly
  aligned with that of the binary, facilitating the settling process
  as the debris orbitally damps. However, this mechanism only works if
  the debris disk can survive long enough to dynamically cool.}

To estimate the production and survival of debris following an impact,
we perform a suite of simulations with Pluto, Charon, and
non-interacting tracer particles.  We select 16 impact sites at random
on the surface of each target and ``eject'' 5,000 tracer particles in
an idealized, hemispherical spray pattern from each site. The ejection
speed of each tracer, $\vej$, defined relative to the target
body, is drawn from a power-law distribution, $f(v>\vej)\sim\vej^{-\alpha}$,
where $\alpha\approx 1$--3
\citep[e.g.,][]{gault:1963, stoff1975, okeefe1985, housen2011}, above
some minimum speed $\vmin$.  Our choice, $\alpha = 2$, suggests
an icy, non-porous target. We expect weaker, porous material to
yield a flatter speed distribution, while a higher tensile strength
steepens it \citep[cf.][see also \citealt{kcb2014}]{svet2011}. When
Charon is the target, we use $\vmin \approx 0.56$~km/s, sufficient to
reach Charon's nominal Hill radius, $\abin (\MC/3\MP)^{1/3} \approx
5.8\RP$\footnote{This value gives roughly the average of Charon's
  distance to the Lagrange points L1 and L2---we do not take into
  account the non-spherical shape of the Roche lobe. The goal here is
  simply to track particle orbits that do not have enough energy to
  reach the satellite zone directly but that might get scattered there
  by the binary.}. When the target is Pluto, $\vmin \approx 1.1$~km/s,
enabling all tracers to reach the nearest point on
Charon's Hill sphere, roughly $10\RP$ from Pluto. With these choices
we track fast particles that travel directly to the satellite
zone, and also those that are scattered into it by close passes with
the binary partners.

With positions and speeds assigned in this way, we evolve the binary
and the tracers forward in time using the sixth-order $n$-body
integrator within our \orchestra\ code \citep{bk2006, bk2011a, kb2008,
  kb2016a}.  For good temporal resolution, time steps range from
15~minutes to less than ten seconds when resolving close encounters
and collisions. This resolution is important even for particles with
semimajor axes in the satellite zone with month-long orbital periods;
their eccentricities tend to be high, and they risk strong and quick
encounters with Pluto or Charon with each pericenter passage.

{The code can also simulate orbital damping, a feature we use to
  estimate the damping rates required to settle tracers on coplanar,
  circular orbits before repeated close encounters with the binary
  remove them \citep{bk2006}. We implement damping by adjusting a
  particle's angular momentum vector, without modification to orbital
  energy or phase.  To shift eccentricity, the code changes the
  magnitude of the angular momentum, deriving new position and
  velocity vectors that preserve the direction of the
  Laplace-Runge-Lenz vector and orbital phase.  The code implements a
  shift in inclination with a coordinate transformation that rotates
  the angular momentum vector toward or away from that of the binary.}

{ We damp tracers only when they are in or beyond the
  satellite zone, since Keplerian orbital elements are not good
  measures of motion close to the binary. This feature is important
  for estimating the damping rates required to settle tracers on
  coplanar, circular orbits before repeated close encounters with the
  binary remove them.}

\begin{figure}[t!]
  \centerline{\includegraphics[width=5in]{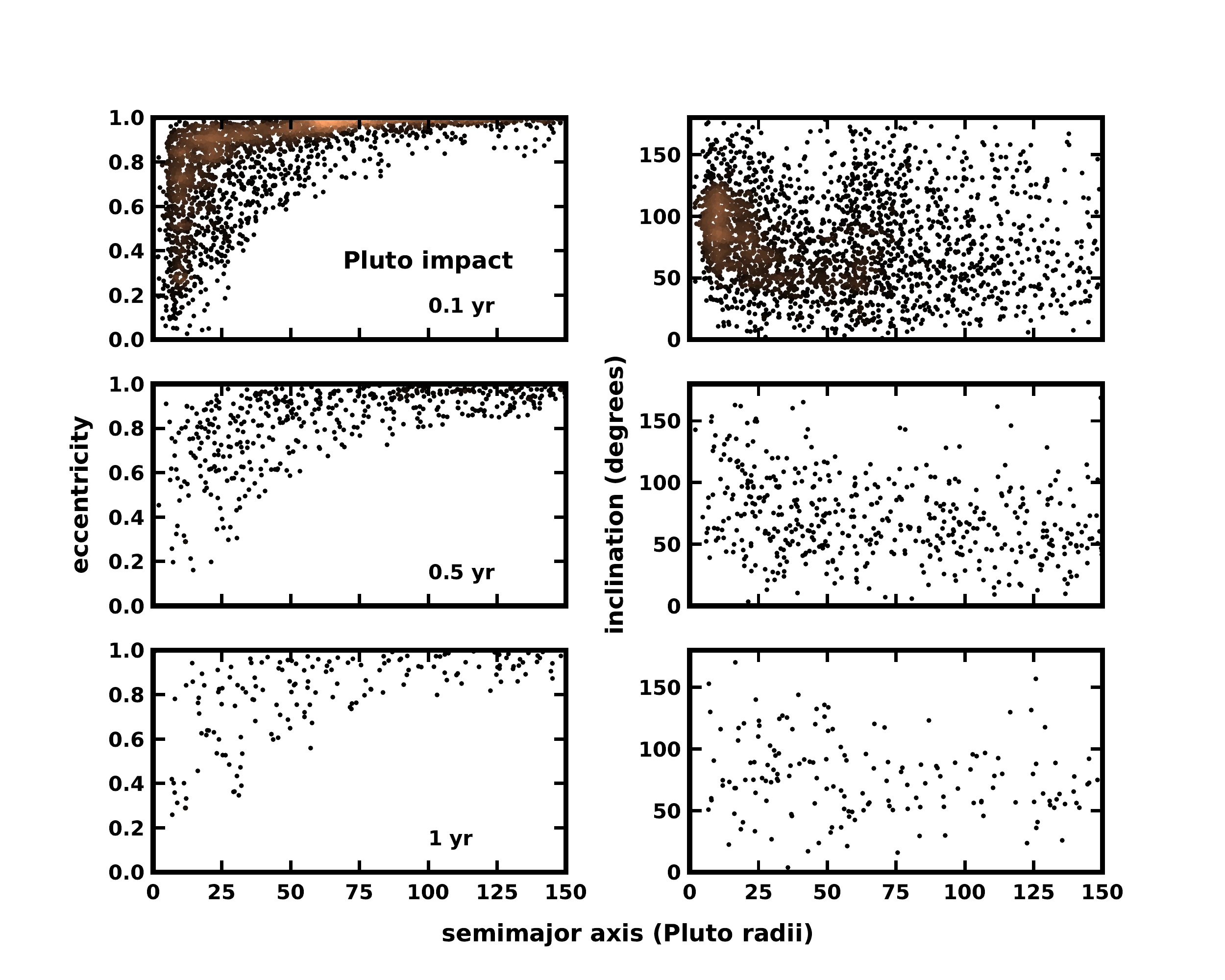}}
  \caption{\label{fig:aiep} The eccentricity and inclination of
    particles representing impact ejecta from Pluto as a function of
    barycentric semimajor axis.  Each row corresponds to a snapshot at
    a specific time after impact, as indicated in the left panels.
    The plots contain tracers from multiple impact events; since the
    adopted spray pattern of ejecta is broad, individual events yield
    similar distributions in this orbital parameter space.  The marker
    color corresponds to local number density of points in each plot
    to help distinguish regions where the markers overlap. For
    comparison, the same color map is used in Fig.~\ref{fig:aiec},
    where the density of points is higher.  The left panels show that
    the eccentricity of bound particles is concentrated at values
    near unity, while the inclination (right panels) is broadly
    distributed.  Repeated encounters with the binary rapidly remove
    particles from the system.}
\end{figure}

\begin{figure}[t!]
  \centerline{\includegraphics[width=5in]{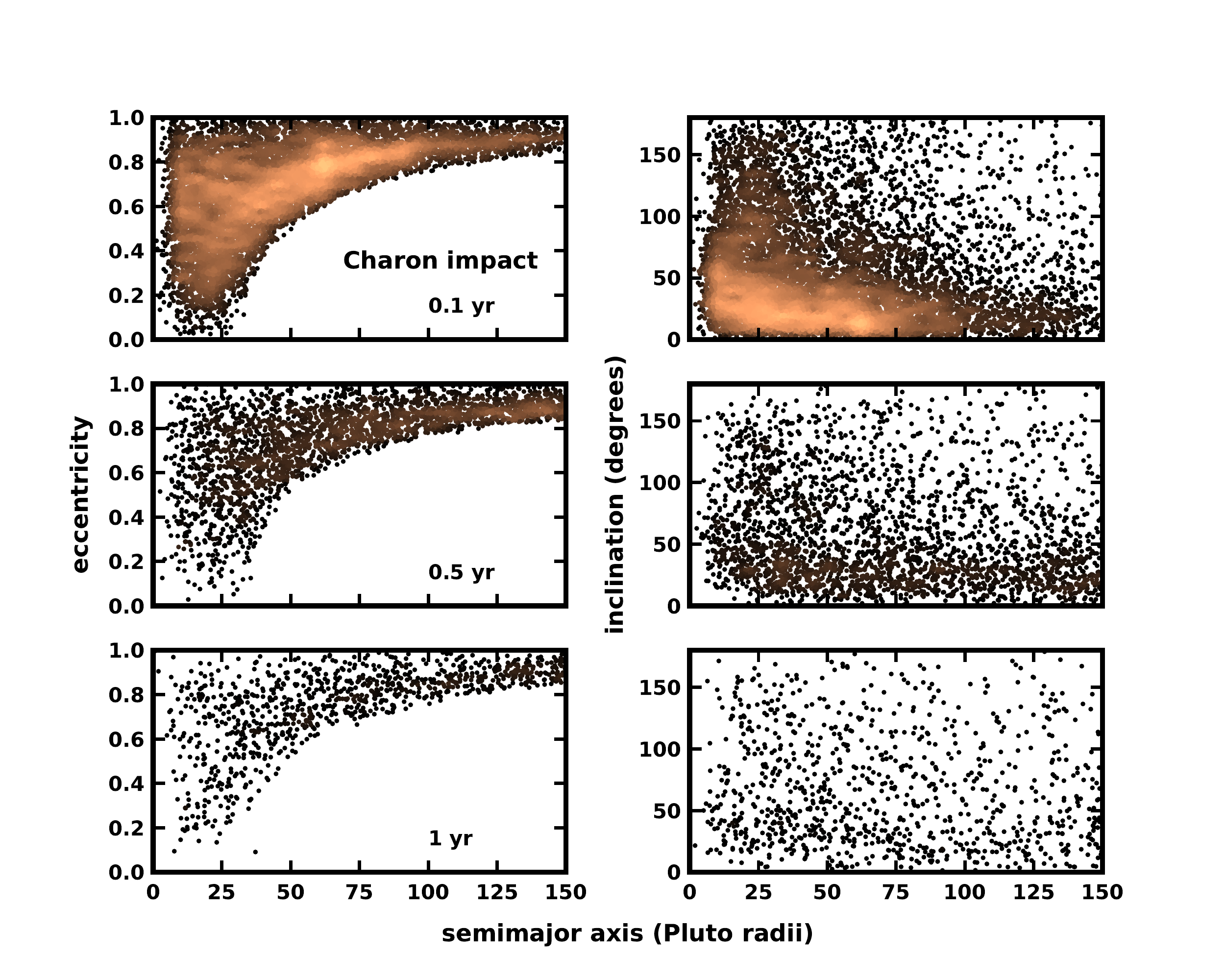}}
  \caption{\label{fig:aiec} The eccentricity and inclination of
    particles representing impact ejecta from Charon, as in
    Figure~\ref{fig:aiep}.  The typical eccentricity of ejecta is
    lower compared with impact ejecta from Pluto, and there is a
    stronger concentration of tracers with low-inclination orbits
    ($\sim$25$^\circ$), showing presence of a thick disk.  Because
    there is no orbital damping for the particles shown here, all will
    eventually get scattered or accreted by the binary.  However,
    since particles in these simulations are launched with a velocity
    boost from Charon's orbital motion, they survive longer than their
    counterparts in Figure~\ref{fig:aiep}.  }
\end{figure}

{We first consider cases with no orbital damping of the ejecta.
  Figures~\ref{fig:aiep} and \ref{fig:aiec} display the results,
  showing the eccentricity and inclination of bound tracer particles
  as a function of semimajor axis at several snapshots in time when
  either Pluto (Fig.~\ref{fig:aiep}) or Charon (Fig.~\ref{fig:aiec})
  is the impact target.  The plots include the ejecta of multiple
  impacts; with a broad spray pattern, individual events yield
  phase-space distributions similar to the composite.  In Figure
  \ref{fig:aiep}, the tracers ejected from Pluto form a roughly
  spherical cloud, with polar orbits more prevalent than
  low-eccentricity, disk-like ones. These orbits are hard to maintain.
  After ejection, tracers fall back close to Pluto, which readily
  scatters or accretes them. Those tracers that additionally lie in
  the plane of the binary may get removed by Charon.}

{In contrast, when debris particles are ejected from Charon, the
  orbits are more closely aligned with the orbital plane of the binary
  (Fig.~\ref{fig:aiec}). The reason is that Charon's orbital motion
  boosts the angular momentum of the ejected particles. The more
  numerous, slower-speed ejecta have specific angular momentum similar
  to Charon itself, yielding a thick, equatorial disk. The scale
  height can be inferred from Figure~\ref{fig:aiec}, which shows a
  concentration of inclinations around 25$^\circ$. Tracers on these
  orbits are significantly longer-lived than their counterparts that
  were ejected from Pluto.  We conclude that an impact on Charon sets
  the stage for the formation of the satellites, while impacts on
  Pluto do not.}

\subsection{Settling into a disk: general requirements}

\begin{figure}[t!]
  \centerline{\includegraphics[width=5in]{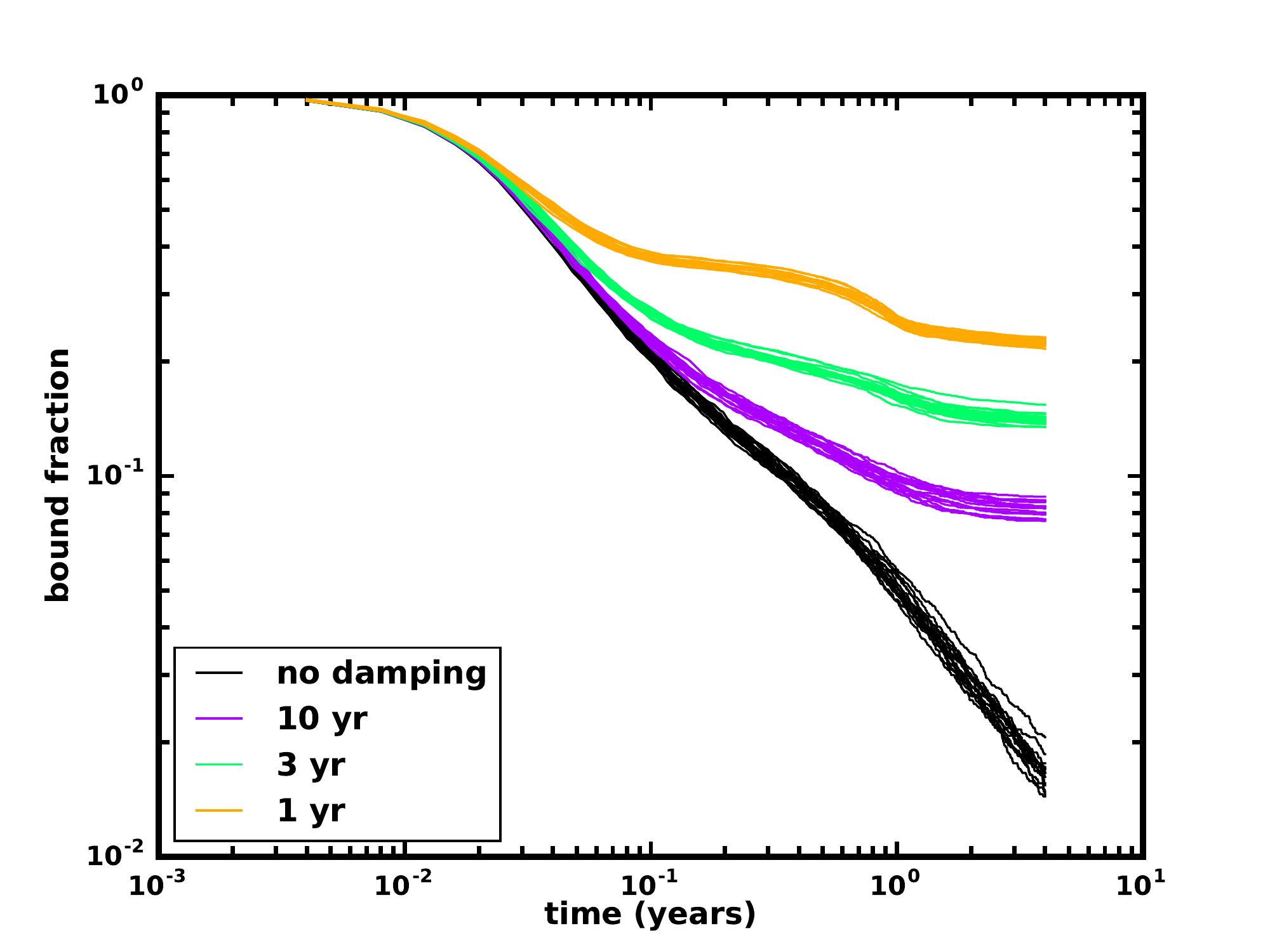}}
\caption{\label{fig:tsurvive} The survival fraction of all simulated
  ejected particles as a function of time since impact. The black
  curves show the fraction of bound tracers for individual impact
  events. The lighter-colored curves correspond to runs with
  eccentricity and inclination damped on time scales as indicated in
  the legend.  When damping times are much longer than a few
  years, less than 10\% of the tracers remain. In all cases, the
  number of particles in the satellite zone is only about 10\% of
  the survivors. Less than a few percent of the total number of
  tracers remain in the satellite zone.
}
\end{figure}

Even when tracers are launched from Charon, the simulations show that
over a period of years, the bound particles return to the vicinity of
the binary time and again until they are either accreted or ejected
(Fig.~\ref{fig:aiec}).  Figure~\ref{fig:tsurvive} emphasizes this
point. The number of bound tracers in the simulations falls steeply in
time; less than two percent of the initial tracer population remains
after four years.  With only the gravity of Pluto and Charon included
in this calculation, we expect that all tracers will eventually be
accreted or ejected.

Despite that bleak result, Figure~\ref{fig:tsurvive} also provides
hope that a reservoir of mass can be preserved to seed the formation
of Pluto-Charon's small satellites.  The plot shows that orbital
damping in eccentricity $e$ and inclination $\imath$ leads to a significant
amount of debris on stable, bound orbits around the binary. For that
demonstration, we adopt a constant time scale $\tau$, damping the
eccentricity and inclination of each tracer at a rate of $e/\tau$ and
$\imath/\tau$, respectively.  Figure~\ref{fig:tsurvive} indicates that a
damping time of no longer than $\tau \sim 5$ years is required to
prevent a sizable fraction ($\gtrsim 10$\%) of the impact ejecta from
being ejected or accreted.

Even in the cases of rapid damping, only $\sim$1\%\ of the total
number of tracers settle in the satellite zone. However, the simple
``in-place'' damping mechanism, where $e$ and $\imath$ are steadily reduced
around the osculating semimajor axis $a$, is not representative of
realistic damping processes. Instead, collisional damping will
concentrate particles where the local number density is high, which
happens to be in the satellite zone (Fig.~\ref{fig:aiec}). If the impact
event generates a cloud of gas, aerodynamic drag can trap small
particles wherever the gas is, even if they would otherwise be on
unbound orbits. Then, as in Figure~\ref{fig:tsurvive}, as much as 20\%
of the material ejected beyond Charon's Hill sphere might wind up in
the satellite zone. We explore these damping mechanisms next.

\subsection{Collisional evolution}\label{sec:collide}

In a swarm of solid particles, collisions drive orbital damping. From
basic kinetic theory, the average rate of collisions experienced by a
particle with radius $\rp$ in a sea of other similar particles scales
as ``$nv\sigma$.''  Applying this estimator to a thick disk of debris
(as in Fig.~\ref{fig:aiec}), the average time between collisions for
each particle is
\begin{eqnarray}\nonumber
  \tcol & \sim &
  \frac{4\pi \rhop \rp^3 a \ieff}{3 \Sigma}
  \cdot
  \frac{1}{\eeff \vkep}
  \cdot
  \frac{1}{4\pi \rp^2}
  \\
    \label{eq:tcol}
  \label{eq:tcolest}
  \ & \sim &
  2.0 \times \rhoponecgs \rptenm
  \Sonecgs^{\!-1}\!
  \afiftyRp^{3/2}
  \itwentyfive
  \eohseven
  \ \text{yr,}  
\end{eqnarray}
where $\rhop$ is the density of solids in the impact debris, $\Sigma$
is the surface density of the debris, $a$ is the semimajor axis,
$\ieff$ is the typical orbital inclination, and $H = a\ieff$ is the
scale height of the debris swarm.  The term $\vkep$ in the denominator
is the local circular speed as if Pluto-Charon were a single point
mass, while  $\eeff$ is an effective eccentricity, so that the product
$\eeff\vkep$ characterizes the random motion of the swarm of impact debris
particles.  Here, we have also used a cross section of $4\pi\rp^2$,
appropriate to a population of identical, hard spheres.

Equation~(\ref{eq:tcolest}) illustrates numerical values of the
collision time for specific parameter choices.  The fiducial value of
$\Sigma = 1$~g/cm$^2$ corresponds to a total mass in the satellite
zone that is comparable to $\Msats$, assuming that the surface density
scales as $\Sigma \sim a^{-1.5}$.  Note that we otherwise do not
consider variation in the surface density of the debris in the above
estimate of the damping time; doing so would steepen the dependence of
$\tcol$ on semimajor axis. For the effective eccentricity, we adopt
$\eeff = 0.7$ on the basis of the tracer simulations. The product
$\eeff \vkep$ with this value is representative of the random speeds
of tracers in the satellite zone.

{Equation~(\ref{eq:tcol}) predicts general trends in the collision
time, such as the decrease in $\tcol$ with increasing number density
of the debris particles. However details of the particle orbits can
affect the numerical estimates.  For example, if all particles were
launched from impact into a narrow eccentric ring, then their relative
velocities would be smaller than the value in
Equation~(\ref{eq:tcol}), resulting in a longer collision
time. Similarly, when debris is launched by an impact on the surface
of a single planet, the bound ejecta will have little spread in
eccentricity as a function of semimajor axis \citep[e.g.,][for a giant
  impact on Mars that may have formed its moons, Phobos and
  Deimos]{hyodo2017a, hyodo2017b}.  However, the ejecta launched from
Charon are relatively large random relative velocities as compared to
an eccentric ring or debris from a single planet \citep[see][Fig.~6,
  therein, compared with Fig.~\ref{fig:aiec} here]{hyodo2017b}.  Thus,
we adopt Equation~(\ref{eq:tcol}) as representative of a debris cloud
with an origin on Charon's surface.}

\subsubsection{Collision outcomes}

The collision time in Equation~(\ref{eq:tcol}) sets the rate that
collisional damping cools a dynamically hot population, causing it to
settle into a thin, coplanar, circumbinary disk. Shortening $\tcol$
increases this rate.  The details depend on the outcomes of myriad
pairwise collisions---the amount of random kinetic energy lost, and
whether particles erode or shatter in the process. The physics of
inelastic collisions \citep[e.g.,][]{porco2008, bk2015pc} and
fragmentation is important. If particles are broken up as the debris
swarm evolves, they become smaller but more numerous, significantly
driving up the collision rate \citep{weth1993, will1994, tanaka1996,
  kl1999a, obrien2003, koba2010a}.

To distinguish collision outcomes, we follow established theory by
comparing the pairwise center-of-mass collision energy $Q_c$ with the
specific binding energy, $\qdstar$ \citep{davis1985, housen1990,
  benz1999, lein2012}:
\begin{equation}\label{eq:qdstar}
  Q_c = \vrel^2/8, \ \ \text{and} \ \
  \qdstar = Q_b\rp^{-0.4}
\end{equation}
where the expression for $Q_c$ applies to equal-mass bodies
\citep{kb2014pc} and $Q_b$ is a material-dependent
strength constant.  This form of $\qdstar$ applies to smaller
bodies for which self-gravity is unimportant \citep[$\rp \lesssim
  0.1$~km, e.g.,][]{leliwa-kopystynski2016}. Particles
are stronger when they are smaller because  they are less
susceptible to internal fractures than their larger counterparts.  We
set $Q_b = 2 \times 10^6$~erg/g$\cdot$cm$^{0.4}$ acknowledging that
literature values range from an order of magnitude lower \citep[``weak
  ice'']{lein2008, lein2009, schlicht2013} to an order of magnitude higher
\citep[``strong ice'']{benz1999} than our choice (``ice'').

By setting $\qdstar = Q_c \sim \vkep^2/8$, we solve for the radius of
particles that marks the transition between disruptive and inelastic
collisions,
\begin{equation}
  \rdstar  =  \left(Q_c/Q_b\right)^{5/2}
  \sim
  0.017 \times \eohseven^{-5} \afiftyRp^{5/2}
  \qbtwoesix^{5/2} \, \text{cm}.
\end{equation}
Particles smaller than $\rdstar$ are able to grow into larger objects
as a result of collisions, while large objects lose mass.

This threshold value of particle radius is sensitive to the semimajor
axis and the assumed particle tensile strength. Weak ice has a
disruption radius of about 0.4~microns. Strong ice has a disruption
radius of a few centimeters.  The threshold radius is even more
sensitive to the effective eccentricity of the debris swarm; $\rdstar$
increases dramatically as $\eeff$ is reduced, since smaller
eccentricities mean lower collision speeds that allow larger particles
to survive.  Thus, velocity evolution is key to the swarm's ultimate
size distribution.

\subsubsection{A simple evolution model}

Guided by detailed simulations showing that a collisional cascade can
produce rapid damping \citep{kb2014pc}, we use an idealized picture to
discern trends in the evolution of the swarm's velocities and size
distribution.  An initial debris swarm consisting of identical
particles experiences a round of collisions in time $\tcol = t_0$,
breaking up all bodies into more numerous smaller ones. This next
generation of debris has both a higher space density and slower random
motions, which together affect the next round of collisions. With each
generation, velocities damp and orbits circularize. In the early
stages when particle sizes and collision speeds are large, disruptive
collisions churn bigger bodies into many smaller ones. Eventually, as
random motions become small, collisions become inelastic with little
fragmentation.  The surviving particles settle into a thin disk on
coplanar, most-circular orbits about the binary planet \citep{lee2006,
  youdin2012, bk2015pc}.

To quantify how velocity and particle size evolve in this scenario, we
assume that with each successive generation of shattering collisions
the radii of particles is a small fraction $\fd$ of their previous
size. A factor $\gd$ gives the corresponding reduction in typical
collision speed (or, equivalently, $\eeff$). Once particle sizes and
velocities are reduced so particles just bounce off each other,
$f\rightarrow 1$, and $\gd \rightarrow\gb$, where $\gb$ is a damping
factor related to the material-dependent coefficients of restitution
\citep[e.g.,][]{bridges1984, sup1995}. Armed with this prescription,
along with the collision time in Equation~(\ref{eq:tcol}), we quantify
the collisional evolution of the system.

If $r_0$ and $e_0$ are the radius and effective eccentricity of the
initial swarm, then after $n$ generations of disruptive collisions,
particles have typical sizes $\fd^n r_0$ and speeds $\gd^n e_0\vkep$.  
The collision time between generations scales as $r/v$; thus,  
\begin{equation}
\tcol = \tcoloh \left(\frac{\fd}{\gd}\right)^{\,n},
\end{equation}
where $\tcoloh$ is the initial, post-impact collision time. 
The number of generations, $n$, if treated formally as a continuous
function of time, is
\begin{equation}\label{eq:ngen}
  n(t) = \text{floor}
  \begin{cases}
    \log\left[1-(1-\fd/\gd) t/\tcoloh\right]/\log(\fd/\gd) & (f\neq \gd)
    \\
    t/\tcoloh & (f = \gd)
  \end{cases}    
\end{equation}
where $t = 0$ is the time of the formation of the debris swarm, and
``floor'' indicates the nearest lower integer.  In general, when the
ratio $\fd/\gd$ is less than one, the collision time gets smaller with
each generation. Otherwise (as is the case for bouncing
collisions), the collision time grows.

What ultimately matters to the survival of the swarm around the
Pluto-Charon binary is the damping time. From our tracer simulations
(e.g., Fig.~\ref{fig:tsurvive}), orbital damping times must be shorter 
than a decade to preserve a substantial reservoir of mass for 
the satellite system.  Here, we define the damping time as
\begin{equation}\label{eq:tdamp}
\tdamp \sim \frac{\eeff}{d\eeff/dt} 
\sim \frac{\ieff}{d\ieff/dt} \sim \tcol / (1-\tilde{g})
\end{equation}
where $\tilde{g}$ is either the velocity factor $\gb$ for bouncing
collisions or $\gd$ for collisions that are disruptive.  While values
of $\gd$ and $\gb$ are uncertain \citep[for examples involving
  coefficients of restitution for water ice, see][and references
  therein]{gartner2017}, we recommend values well below $0.5$. This
range acknowledges that in high-speed disruptive collisions,
much of the kinetic energy is dissipated as heat.  In bouncing collisions,
heat, compactification, and cratering damp rebound speeds
\citep[cf.][Fig.~22 therein]{porco2008}.  Thus, $\tdamp$ is 1--2 times
$\tcol$. Within a few generations of collisions, orbits are fairly
circular and the debris is safe from close encounters with the central
binary.  We consider specific scenarios next.

\subsubsection{Large debris particles, disruptive collisions}

If debris particles are initially large $(\rp \gg \rdstar)$, the first
generations of collisions grind them down to small sizes.  From
Equation~(\ref{eq:ngen}), the number of generations required until 
shattering stops and the swarm particles reach their final size is:
\begin{equation}
  n_{cc} \sim \left[ 11 + \log(r_0/10~\text{m}) + 5\log(e_0/0.7) - (5/2) 
\log(a/50\RP \cdot
    Q_b/Q_{b}^\prime) \right]
  / \left|\log(\fd) + 5 \log(\gd)\right|,
\end{equation}
where $Q_{b}^\prime$ is our fiducial value ($2\times 10^6$~[cgs]), and
we have assumed that both $f$ and $\gd$ are less than or equal to one;
the subscript ``cc'' is a reference to ``collisional cascade.''
Setting $\fd = \gd = 0.5$ for a swarm of particles with $r_0 = 10$~m,
it takes just three generations for the eccentricity to fall below 0.1
and for the particles to reach their final size of about 1~m.  Still,
the damping time is twice the collision time, so that with a mass
$\Msats$ in the satellite zone (as in Eq.~(\ref{eq:tcol})), three
generations takes about a decade. As in Figure~\ref{fig:tsurvive},
more than 90\%\ of the debris will be lost to interactions with the
binary in this time.  A substantially more massive disk
($\gtrsim$10$\Msats$ in the satellite zone), a smaller initial radius
($<$10~m), or (as we expect) lower values of $\fd$ and $\gd$ would
help to speed up damping and preserve more material in the satellite
zone.

\subsubsection{Small debris particles, bouncing collisions}

Mutual collisions between sub-millimeter particles are inelastic,
resulting in reduced random speeds and little change in particle
sizes. Because particles are small and their number density is high
even when the satellite zone has only the minimum mass needed to
account for the satellites, collision times are very short. For
example, a swarm of strong ice, made of ``indestructible'' particles
as large as a few centimeters, has a collision time of less than a
day. Even if coefficients of restitution are large ($\sim$0.5), the
time to damp from $\eeff \approx 0.7$ to less than 10\% of that value
is months, not years. In general we expect a swarm of small debris
particles to damp and circularize quickly and efficiently.

Small particles, particularly the sub-micron grains of radii $\rdstar$
for weak ice, may be lost to solar wind and radiation pressure
\citep[e.g.,][]{pires2013, gaslacgallardo2019}. A debris swarm is protected
from these and other effects if it is optically thick. We measure the
optical depth of debris in the satellite zone in the vertical and
radial directions, along rays that go through the plane of the thick
disk, and along a path from the binary center of mass outward through
the disk's equatorial plane:
\begin{eqnarray}
  \tau_z & = & \frac{3\Sigma}{4\rhop\rp}
  \sim  0.75 \times
  \Sonecgs\rponecm^{\!-1}\rhoponecgs^{\!-1} \ \text{yr}
  \\
\tau_r & = & \frac{3 \Dasatzone\,\Sigma}{4\rhop \rp\, a \ieff} 
 \sim 
1.1 \times \Dathirtythree \Sonecgs
\rponecm^{\!-1} \rhoponecgs^{\!-1} \afiftyRp^{\!-1}
\itwentyfive^{\!-1}  \ \text{yr}
\end{eqnarray}
Particles smaller than roughly a centimeter thus form an optically
thick cloud, conferring protection from significant losses from
solar radiation.

We conclude that collisional damping, accelerated by a collisional
cascade, is an effective means to rapidly damp a swarm of ejecta from
a TNO-Charon impact, provided that the debris particles are numerous
and small enough. If the typical debris particle less than
$\sim$10~meters in radius, and if there is more than {$\sim$}$\Msats$
worth of debris in the satellite zone, collisional damping leads to
the formation of a circumbinary disk at the right location for
building the satellites.

\subsection{Damping in a gas cloud}\label{sec:gas}

If a circumbinary gas cloud forms during an impact, gas drag may be 
a source of orbital damping. We explore how such a cloud might affect
the orbital dynamics of the solid debris.  First we consider
preliminaries. The sound speed in the gas is
\begin{equation}
  \cs = \sqrt{\frac{\gammaadiabat kT}{\mumH}} \approx
  0.15 \Tforty^{1/2} \mumHwater^{-1/2}
  ~\text{km/s}
\end{equation}
where we let $\gammaadiabat = 1.3$ and molecular weight $\mumolec =
18$, corresponding to water. This speed is comparable to the orbital
speed in the middle of the satellite zone, $\vkep \approx 0.13$~km/s
at $50\RP$. The mean thermal speed is larger, $\vtherm \approx
1.4\cs$. We choose our fiducial temperature, $T = 40$~K,
to be roughly the equilibrium temperature at Pluto-Charon's orbital
distance from the Sun.

The temperature of the gas cloud is coupled with the cloud's structure
and fate around the central binary.  The thickness of a gas disk
around a central mass scales as $\cs/\vkep$ times the orbital distance
\citep[e.g.,][]{ss1973, lbp1974}.  Even at $T = 40$~K, and despite the
angular momentum imparted to it from the Charon impact, the gas is too
hot to settle into a disk.  The sound speed is also close to the
escape speed, $\vesc \approx 0.18$~km/s at $a=50 \RP$, indicating that
the gas cloud will evaporate within a few dynamical times.  Thus, we
picture a scenario where the gas forms a short-lived, {roughly
spherical,} rotating cloud.

Assigning a fiducial mass in gas of $\Mgas = 10^{19}$~g, distributed
uniformly in a spherical shell that spans the satellite zone,
the mean free path of gas molecules is 
\begin{equation}
  \mfp \approx \frac{\mumH}{\pi\rmolec^2 \rhogas}
 \approx  24 \mumHwater \Mgasnineteen^{-1} \rmolecNtwo^{-2}  ~\text{m}.
\end{equation}
This parameter helps to set how solid bodies interact with the
gas. When a small debris particle of radius $\rp \lesssim \mfp$ has a
speed $\vprel$ relative to the gas, the drag force decelerates the
particle at a rate of
\begin{eqnarray}\label{eq:adragmy}
  \adrag & \approx & \frac{C_{\rm Eps}\rhogas}{\rhop\rp}
  \left(\vtherm + \vprel/4\right)
  \vprel
\end{eqnarray}
where the constant $C_{\rm Eps} = 1$ for specular reflection and is
generally near unity for elastic collisions \citep{whipple1972,
  ada76, weiden1977a, raf2004}.  In the limit of slow particle speed
($\vprel \ll \vtherm$), this expression is the Epstein drag law. The
high speed limit ($\vprel \gg \vtherm$) derives from a simple
ballistic approximation where all molecules have a constant velocity
in the frame of the macroscopic body. Interpolation between these two
extremes gives us an estimate for cases relevant here, with $\vprel
\sim \vtherm$.

In a tenuous gas with $\vprel \approx \vtherm \approx v_K$, the
magnitude of the drag acceleration of a debris particle in the
satellite zone scales roughly as $\vkep^2$.  Thus, the ``stopping
time,'' which characterizes the impact of gas drag on the dynamics of
a solid particle, is
\begin{eqnarray}\nonumber
  \tstop & \equiv & \frac{\vp}{\adrag} \sim \frac{\rhop\rp}{\rhogas\vprel}
  \\ \ & \sim &
  16 \rhoponecgs \rponemm \Mgasnineteen^{\!-1} \afiftyRp^{3/2} \ \text{days}.
\end{eqnarray}
The length of time in the lower equation is comparable to the dynamical
time of orbits in the satellite zone. If the gas cloud is 10\%\ of the 
mass of the present day satellites, then particles much smaller than 1~mm 
are entrained in the gas. Much larger objects do not notice the cloud.

{ From these results, there may be a pathway for the formation of
  a debris disk in the satellite zone. If the conditions are right for
  a collisional cascade, then much of the impact debris will be
  quickly converted into sub-millimeter grains and entrained in a
  rotating gas cloud.  As the cloud evaporates and particles settle to
  the midplane, the solids form a thin disk with enough mass to move
  independently of the gas.  Once the cloud vanishes, the remaining
  solids are free to coagulate and grow, as in a circumstellar
  planetary system. }

 {However, the plausibility of this scenario depends on a number
   of uncertain factors, including the way the gas orbits the
   binary and how it disperses. For example, if the gas rotation is
   much slower than the local Keplerian speed, entrained solids will
   simply fall into the binary after the gas disperses.  Otherwise,
   the presence of a cloud might help retain solid material that would
   have been removed by interactions with the binary. }

\subsection{Trapping mass in the satellite zone}\label{subsec:trap}

{ In our tracer simulations of an impact on Charon, orbital damping
saves as much as 20\%\ of debris particles from ejection or accretion
by the binary. Yet precious little of the
ejecta ends up in the satellite zone when particles damp ``in
place'', even if they do so very quickly. Fortunately,
significantly more mass likely
ends up in the satellite zone than settles there in our tracer simulations.
After a few binary orbits, about 3\% of the total number of tracers
have semimajor axes in the satellite zone, and over
10\% have orbits that will pass into it.  Collisions between these
objects, along with others that are scattered into the satellite zone
at different times, can trap material there.  In this way, the mass 
in the satellite zone can be increased substantially.}

{ Ejected debris crossing into the satellite zone will probably
  settle there if it participates in the collisional cascade at all.
  Solids orbiting interior to the satellite zone have short collision
  times and may be part of a robust collisional cascade closer to the
  binary (Eq.~(\ref{eq:tcol})), but the strong dynamical excitation by
  the binary prevents settling. Instead, eccentricity pumping pushes
  debris particles into the satellite zone where they are swept up by collisions
  with other debris there. While these trapping mechanisms will have
  some impact on the distribution of mass and angular momentum of
  stable material within the satellite zone, we anticipate that they
  are effective in trapping virtually all of the small debris
  particles that stray into the satellite zone.  Thus, as much as
  10\%\ of the impact debris ejected beyond Charon's Hill sphere, can
  wind up in the satellite zone as mass available to build Hydra and
  company.}

If trapping is necessary to increase the mass in the satellite zone
to equal $\Msats$, then the bulk of the debris particles must start off
small, with a collision rate high enough to trigger the cascade.
Assuming that the satellite zone can trap roughly three times the mass
that it contains just after ejecta reaches it, then it might start out
with a mass of only $0.3\Msats$. Then, from Equation~(\ref{eq:tcol}),
the starting radius must be no bigger than a few meters.  In a debris
swarm like this, the collisional cascade and trapping conspire to triple
the mass in the satellite zone, delivering exactly the right amount
for the satellites.

Many of the larger bodies that are not caught up in the collisional
cascade pass into or completely through the satellite zone and may get
trapped by the smaller debris that is forming a disk there.  Trapping
mechanisms include (i) dynamical friction between large impact
fragments and gas \citep{ostriker1999} or small debris
\citep[e.g.,][]{gold2004}, (ii) kinematic friction caused by
collisions with the small solids already settled in the satellite
zone, and (iii) erosion of an interloper by the small debris. From
analytical estimates \citep[e.g.,][]{ostriker1999, bk2014}, the first
mechanism---dynamical friction---is too slow-acting to be relevant in
a scenario where rapid damping is essential. Thus we focus on
kinematic friction and erosion.

To explore these possibilities, we assume that the satellite zone
contains a disk of small ($<$1~cm) particles with a total mass of
$10^{20}$~g, spread uniformly across the disk with a typical
inclination of $\ieff = 5^\circ$.  A large body, with pericenter close
to the binary, apocenter well beyond the satellite zone, and an
inclination that is also around $\ieff$, passes through the satellite zone on
a nearly radial path. As it does so, it experiences a loss of orbital
energy given by the ballistic limit of Equation~(\ref{eq:adragmy}).
{ By substituting gas density with that of the debris, and ignoring the
  debris particles' thermal (random) motion, the energy change is}
\begin{equation}\label{eq:acolmy}
\Delta E \sim -\frac{\vPrel^2}{\rhoP \rP}\frac{\Dasatzone \Sigma}{a\ieff},
\end{equation}
where $\rP$, $\rhoP$, and $\vPrel$ all refer to the large body.
This loss of energy translates into a change of semimajor axis,
\begin{equation}
  \frac{d\aP}{dt} \sim 2\times \frac{\Delta E}{\pi} \sqrt{\frac{\aP}{GM}},
\end{equation}
where the factor of two accounts for energy loss during 
both inward and outward passages through the satellite zone.
An estimate of the orbital decay time is then
\begin{eqnarray}\nonumber
  \tdecay & = & \frac{\aP}{d\aP/dt} \approx
  \frac{\pi}{2} \frac{\rhoP\rP}{\Sigma} \frac{a\ieff}{\Dasatzone}\frac{\sqrt{GM\aP}}{\vrel^2} 
   \\ \label{eq:tdecayfar}
  \ & \approx &
  4.3 \times
  \rhoPonecgs \rPtenm 
  \Sonecgs^{\!-1} \Dathirtythree^{\!-1} \afiftyRp^{2} \ifive \ 
  \aPonehundredRp^{1/2}
  \left[\frac{\vPrel}{\vkep}\right]^2
\text{yr},
\end{eqnarray}
where we assume that the speed of the body relative to the swarm is
approximately the local circular speed in the satellite zone. If the
interloping body has a higher inclination that $\ieff$, the decay time
is substantially longer. A body on an eccentric orbit that lies well
out of the plane of the disk will not interact with the debris swarm
at all. 

When a large impact fragment is on a close-in orbit that extends into
the satellite zone but not beyond it, collisions with the
faster-moving small debris there raise its semimajor axis and lower
its eccentricity.  After repeated excursions, the object eventually
circularizes in the satellite zone.  The energy change per orbit,
$\Delta E$, is characterized by the acceleration from collisions
(Eq.~(\ref{eq:adragmy})) times the path length of the interloper's
excursion into the satellite zone.  Picturing a body that has a
semimajor axis $\aP$ near the inner edge of the satellite zone (not
deep in the unstable region) and sufficient eccentricity $\eP$ to
reach the middle of the zone, we estimate
\begin{equation}
\Delta E \sim +\frac{\pi \vPrel^2}{4 \rhoP \rP} \frac{(\aP+a)\Sigma}{4 a\ieff}.
\end{equation}
{ since collisions within the stream of
  small particles boost the larger body's
  speed, providing the torque needed to circularizes it.}

The time scale for boosting the object from a low-$\aP$,
high-$\eP$ orbit interior to the satellite zone to a circular orbit in
the zone is
\begin{eqnarray}\nonumber
  \tboost & = & \frac{\aP}{d\aP/dt} \approx
  4\frac{\rhoP\rP}{\Sigma}\frac{a^2\ieff (1+\eP/2)}{\sqrt{GM}\aP^{3/2}}
   \\ \label{eq:tdecaynear}
  \ & \approx &
  5.3 \times \left[\opefac\right]
  \rhoPonecgs \rPtenm 
  \Sonecgs^{\!-1} 
  \aPthirtyRp^{-1/2} 
  \afiftyRp^{2} \ifive \ \text{yr}.
\end{eqnarray}
This estimate illustrates that mass can also move from the unstable region
into the satellite zone.

Unless the mass in the debris is significantly larger than the present
day satellite-zone mass, the maximum size of particles trapped in this
way is $\sim$10~m. This value is disappointingly small if we hope for
kilometer-size seeds or ready-made satellites the size of Styx or Kerberos.
Furthermore, because of the geometry of the disk, trapping is
efficient only for the fraction of 10-meter particles with low
inclination. Still, the impact debris is generally launched from
Charon with an effective inclination of $\sim$25$^\circ$ so that a
substantial fraction these larger bodies has an inclination that is
low enough to interact with the debris swarm within the
satellite zone (Fig.~\ref{fig:aiec}).

{ In addition to orbital damping, larger fragments can also be
  ground down by cratering collisions with the swarm of smaller
  particles. If a large body plunges through the satellite zone from
  well outside it on an eccentric, low-inclination orbit,
  ($\imath\sim\ieff$), it collides with many smaller particles, eroding its
  surface. As seen in laboratory experiments \citep[Fig.~16
    therein]{housen2011} and simulations \citep[e.g.,][]{svet2011},
  each high-speed collision ($\vrel \sim \vkep$) can eject much more
  than the mass of the small impactor.  By assuming that a debris
  particle removes at least its own mass from the fragment after each
  collision, then the fragment's fractional mass loss per orbit is}
\begin{equation}\label{eq:erode}
  \frac{\Delta{\mP}}{\mP} \gtrsim \frac{3}{2\rhoP\rP}
  \frac{\Dasatzone\Sigma}{a\ieff}
  \sim 0.011 \rhoPonecgs^{\!-1}\rPtenm^{\!-1}
  \Dathirtythree\Sonecgs\afiftyRp^{\!-1}\ifive^{\!-1},
\end{equation}
where $\mP$ is the mass of the large body, { and we use its geometric
cross section to derive the debris mass that it encounters}.  From the
rightmost expression, it is clear that a 10-meter fragment on an
eccentric orbit with a semimajor axis of $100\RP$ loses almost 5\%\ of
its mass per year. Smaller particles lose mass at even higher rates.

While a small fragment loses a greater fraction of its mass to erosion
than a large fragment, the larger body loses more mass in absolute
terms. From Equation~(\ref{eq:erode}), a kilometer-size body, with
mass $4\times 10^{15}~g$, loses $5\times 10^{11}$~g per orbit, a
thousand times more than a 10-meter interloper. Still, as a way to add
mass to the satellite zone, the demographics favor the smaller bodies:
The number of erosive collisions involving 10-meter objects is much
larger than for 1-kilometer bodies, and the amount of mass delivered
by the smaller bodies is greater as well.  If there were as
much mass in 10-meter objects passing through the satellite zone as
there were small debris in the disk, then the disk mass would increase
by 25\% in five years, depending on the typical inclination of the
interloping bodies. Smaller interlopers would contribute even more.

\subsection{Summary}

From this analysis, we conclude that ejecta from a major impact on
Charon can be trapped within the satellite zone. Our main results are
(i) the delivery of material to the satellite zone for the formation
of the satellites requires a TNO impact with Charon, and (ii) the
impact ejecta must consist of debris particles with radii of less than
about 10~meters to orbitally damp to form a circumbinary disk
in the satellite zone. If the material strength of the debris is not
strong enough to prevent shattering, a collisional cascade is
triggered, yielding a dynamically cool circumbinary disk composed of 
particles no bigger than a few centimeters in radius. In contrast, 
objects larger than 10 meters are likely lost to accretion or ejection 
by the Pluto-Charon binary.
%


{
Collisional damping is at the heart of these impact scenarios. The
success of any model requires robust collisional damping to produce a
reservoir of circumbinary mass for the satellites.  Only debris that
can participate in this damping is relevant. All else is lost.
}

When the evolution of debris from a direct TNO impact on Charon is driven by
collisions, the delivery of material to the satellite zone is not
efficient.  Only about 10\%\ of the high-speed ejecta dynamically
settles in the satellite zone.  Higher efficiency may be achieved in
other scenarios, as when the debris is entirely composed of
micron-size grains that behave like a viscous medium, or if there were
a transient gas cloud to aerodynamically trap small debris
particles. Another possibility is if a surface-skimming impact
concentrates material on orbits that fortuitously lie in the plane of
the binary with semimajor axes within the satellite zone. Our simple
model of debris from of a direct hit on Charon nonetheless provides a
baseline for the successful delivery of the building blocks for the
satellites from a TNO impact.

\section{A Full Demonstration with the ORCHESTRA code}\label{sec:orch}

\subsection{Initial Conditions}

To explore this model in more detail, we perform several calculations
with the full \orch\ code.  In the multiannulus coagulation routine, an 
area between 19~\RP\ and 135~\RP\
contains 28 concentric annuli distributed in equal intervals of
$a^{1/2}$. Each annulus has 80 mass bins with minimum radius $r_{\rm
  min} = 0.01$~$\mu$m\ and maximum radius $r_{\rm max} = 1$~m. We
seed this grid with solids having total mass $M_0 = 10^{20}$~g,
material density 1.5~g~cm$^{-3}$, surface density $\Sigma \propto
a^{-2}$, and a size distribution $n(r) \propto r^{-3.5}$.  With these
initial conditions, most of the mass lies in the largest objects;
annuli in the inner portion of the grid have roughly twice the total
mass each annulus in the outer portion.

For the eccentricity and inclination of the solids, we consider three
sets of initial conditions that span likely outcomes for material
ejected from a collision between a KBO and either Charon or
Pluto. Solids lie within a thick disk with opening angle $\imath_0$ =
15$^\circ$; the initial vertical velocity ranges from $v_z \approx$
35~\ms\ at the inner edge of the disk to $v_z \approx$ 14~\ms\ at the
outer edge of the disk.  For the initial eccentricity $e_0$, we assume
orbits with pericenters (a) near the orbit of Charon ($q_a$ = 15~\RP;
$e_0$ = 0.25 at 20~\RP\ and $e_0$ = 0.89 at 135~\RP), (b) midway
between the orbits of Charon and Pluto ($q_b$ = 10~\RP, $e_0$ = 0.5 at
20~\RP\ and $e_0$ = 0.925 at 135~\RP), and (c) near the orbit of Pluto
($q_c$ = 2~\RP; $e_0$ = 0.9 at 20~\RP\ and $e_0$ = 0.985 at 135~\RP).

To allow the $e$ and $\imath$ of mass bins to react to the gravity of Pluto
and Charon, we assign massless tracer particles to each mass
bin. Within an annulus, each mass bin is allocated 160 tracers (12800
tracers for 80 mass bins). Another 3200 tracers are assigned to bins
with the most mass per bin. The number (mass) of particles assigned to
each tracer is based on the total number (mass) of particles in a mass
bin divided by the number of tracers assigned to that bin. In mass
bins with few particles, the algorithm assigns integer numbers of
particles to each tracer. Thus, some tracers may `carry' more mass
than other tracers.  Orbital $e_0$ and $\imath_0$ for each tracer
follow the $e_0$ and $\imath_0$ for the assigned mass bin. Tracers
initially have random orbital phases; some tracers are initially near
orbital pericenter, while others are near apocenter. Although the
midplane of the disk lies in the Pluto--Charon orbital plane, some
tracers initially lie within the orbital plane while others begin
their evolution out of the orbital plane.  Once assigned to a mass
bin, a tracer may move to another annulus in response to gravitational
interactions with Pluto, Charon, and the mass in the coagulation grid,
but it may not move among mass bins.

In these examples, we do not assign tracers to mass bins with no mass
initially. As the calculation proceeds, catastrophic disruption and
cratering remove mass from the higher mass bins; fragments of these
collisions are placed in lower mass bins. Thus, the largest objects do
not grow from mergers; higher mass bins remain empty for the duration
of the calculation. In future studies, we plan to investigate the
growth of particles that remain in the grid once the damping of
relative velocities allows collisions to create larger merged objects.

\subsection{Calculational Approach}

From this initial setup, calculations proceed as follows. With a time
step of length $\Delta t$, the coagulation code derives the changes in
the number, total mass, $e$, and $\imath$ of each mass bin from
collisions and orbital interactions with all other mass
bins. Collision rates are derived from the particle-in-a-box algorithm
\citep[e.g.,][]{kb2002a,kb2004a,kb2008}. For each mass bin $k$ in
annulus $i$, the rate of collisions with solids in mass bin $l$ in
annulus $j$ is a function of the number density of particles ($n_{ik}$
and $n_{jl}$), the collision cross-section, the relative velocity, and
the overlap of orbits \citep[see sec.~2 of][]{kb2008}.  When $i = j$,
the overlap is 1; otherwise, the overlap is approximately the ratio of
the volume of annulus $j$ that lies within the volume of annulus $i$.

For the initial particle velocities considered here, all collisions
are destructive, with outcomes set by the ratio of the center-of-mass
collision energy $Q_c$ to $\qdstar$.  The mass ejected in a collision is
$m_e = (m_1 + m_2) (Q_c / \qdstar)$ where $m_1$ and $m_2$ are the
masses of the colliding planetesimals. Within the ejecta, the largest
object has a mass $m_l = m_e (\qdstar / Q_C)^{b_l}$, where $b_l$ = 1.
Smaller fragments follow a power-law size distribution, $n(r) \propto
r^{-q}$ with $q$ = 3.5.  Other choices for $b_l$ and $q$ have little
impact on the evolution \citep{kb2016a,kb2017a}.  With no large
objects in the grid, collisional damping dominates dynamical friction
and viscous stirring. The Fokker-Planck algorithm within \orch\ solves
for the damping of each mass bin \citep{oht1992,oht2002,kb2008}.

At the end of the coagulation step, each tracer is assigned a target
eccentricity $e_t$ and inclination $\imath_t$ based on its current
$e_i$ and $\imath_i$ and the change in $e$ and $\imath$ for its mass
bin from the coagulation calculation. These targets result in time
derivatives for $e$ and $\imath$, $de/dt = (e_t - e_i) / \Delta t$ and
$d\imath/dt = (\imath_t - \imath_i) / \Delta t$.  Before the
\nbody\ step, tracers are also assigned new numbers and masses of
particles based on the number and mass within each mass bin. Although
the \nbody\ code does not use this information, each tracer carries a
changing mass of solids based on the evolution of solids in the
coagulation grid.

Within the \nbody\ step, algorithms update the positions and
velocities of tracers and Pluto--Charon.  Tracers evolve with their
derived $de/dt$ and $d\imath/dt$ and respond to gravitational
interactions with Pluto and Charon. The orbits are evolved with a
sixth-order symplectic integrator with 200 steps per binary orbit.  The
tracers' eccentricity $e$ and inclination $\imath$ are shifted
incrementally over time steps set by the coagulation code, with shift
rates determined according to the damping and stirring inferred from
the coagulation calculations.  These changes to $e$ and $\imath$ are
implemented by small adjustments to the direction of travel and (if
necessary) incremental shifts in position, but without affecting the
other osculating orbital elements. 

At the end of the \nbody\ step, tracers have new positions,
velocities, and orbital elements $a$, $e$, and $\imath$. Tracers with
new $a$ substantially different from the `old' $a$ are placed in new
annuli. When a tracer lands in a new annulus, the number and mass of
particles assigned to that tracer move out of the mass bin in the old
annulus and into a mass bin within the new annulus. Some tracers
collide with Pluto or Charon; others are ejected beyond the outer
limits of the \nbody\ calculation space, $a \gtrsim 1000~\RP$. After
these tracers are deactivated for the remainder of the calculation,
the coagulation particles associated with these tracers are removed
from the coagulation grid.  When an active tracer has a semimajor axis
outside of the coagulation grid, its mass is removed from the old
annulus and is not placed in a new annulus.  If that tracer returns to
the grid before a collision with Pluto--Charon or ejection, the mass
that it carries also returns to the grid.

Although assigning tracers to annuli based on their current position
$(x, y, z)$ seems reasonable, placement based on $a$ is more in the
spirit of the coagulation code.  The collision and Fokker-Planck
algorithms derive rates based on particle volumes, $V = 4 \pi a \Delta
a H$, where $\Delta a \approx \delta a + e a$ and $\delta a$ is the
physical width of the annulus \citep{kb2008}. In this application, the
physical extent of particle orbits is much larger than the physical
width of each annulus. Thus, placing tracers in annuli according to
$a$ recovers the correct volume for calculations of collision and
stirring rates.

The new orbital elements for tracers inform the $e$ and $\imath$ of
mass bins in the coagulation code. Within each mass bin, we derive the
median $e$ and $\imath$ and their inter-quartile ranges for the set of
tracers assigned to that mass bin. These medians set the new $e$ and
$\imath$ for the mass bin. The inter-quartile ranges allow us to
monitor the accuracy of the median in measuring the typical $e$ and
$\imath$ for a set of tracers. Typically, the inter-quartile ranges
are small.

This set of steps provides a closed-loop algorithm that allows the
mass bins and the tracers to respond to the gravity of Pluto--Charon
and the orbiting solids.  The coagulation particles tell the tracers
how to react to solid material orbiting Pluto--Charon.  In turn, the
tracers tell the solid material how to react to Pluto--Charon. As long
as time steps are not too long, the lag between the coagulation and
\nbody\ steps does not introduce significant offsets in the evolution
of the mass bins and the tracer particles.  Aside from setting the
length of time steps based on the changing properties of the mass bins
and the accuracy of the \nbody\ integrator, the code has several
constraints to make sure that changes in the evolution of tracer
orbital elements within the \nbody\ code are well-matched to changes
in the evolution of the mass bins.

Although this algorithm follows the evolution of $e$ and $\imath$
well, it does not include a mechanism to transfer tracers from one
annulus to another due to a catastrophic collision.  In the
coagulation code, collision of particles in annulus $i$ in another
annulus $j$ results in debris deposited in an intermediate annulus
that allows conservation of angular momentum, $j^\prime \approx (i +
j) / 2$.  Formally, we could derive a $da/dt$ to be applied to tracers
in the \nbody\ code that would transport the appropriate tracers to
the `correct' semimajor axis. However, this approach is
computationally expensive. For the present study, we allow the
coagulation code to transfer material from one annulus to another
through cratering collisions and catastrophic disruptions.  Active
tracers in annuli that receive additional material from collisional
evolution are assigned more or less mass every time step based on the
changing contents of mass bins in the coagulation code.

\subsection{Results}

The evolution of the orbiting solids in calculations with different
$q$ ($q_a, q_b, q_c = 15, 10, 2~\RP$) all follow a similar path.
Prograde orbits with $a \lesssim 1.7 a_{PC}$ are unstable
\citep{kb2019a}. Despite short collisional damping time scales for
solids with $a \approx$ 18--30~\RP, the central binary gradually
clears away tracers and their associated solids in this semimajor axis
range on one year time scales. Before this clearing is complete,
destructive collisions between particles in the innermost annuli and
other mass bins deposit debris in other annuli. This process
effectively transports some material from inside the unstable region
to annuli outside the unstable region. As tracers in the inner disk
are damped and cleared, some move onto orbits with larger $a$ (due to
dynamical interactions with Pluto--Charon) but lower $e$ and $\imath$
(due to damping). Over time, these tracers and their associated solids
may remain on stable orbits in the outer part of the disk.

At large distances, $a \approx$ 100--135~\RP, collision time scales
are 40--50 times longer than time scales at 30--40~\RP. Solids in this
semimajor axis range have $e \approx$ 0.9 and damp slowly. Although
collisions deposit debris into the inner disk, transport of material
from the outer disk inward is much slower than transport from the
inner disk outward. As this material evolves, tracers on high $e$
orbits often interact with Charon. The slow progress of collisional
damping prevents these tracers from evolving to smaller $e$ before
they suffer a large impulsive encounter with Charon.  Over time, the
central binary ejects nearly all of this material.

At intermediate distances from the central binary, $a \approx$
30--100~\RP, the fate of solids depends on the initial
eccentricity. In systems with $q \approx q_c - q_b$, collisional
damping needs to raise the pericenter to $q \gtrsim$ 25~\RP\ to avoid
strong dynamical encounters with the central binary. Although damping
raises $q$, this evolution is slow compared to the loss of material
from dynamical encounters with Pluto or Charon. In model (b), it takes
1~yr (10~yr) to remove 93\% (98\%) of the solids.  Model (c) evolves
faster: it takes less than a day (month) to lose more than 90\% (99\%)
of the solids. After 100 yr, both models settle into a steady-state,
where the mass changes very slowly with time. At this point, model (b)
(model(c)) has 2\% (0.006\%) of its initial mass remaining in orbit
around Pluto--Charon.

When $q \approx q_a$, collisional damping works fast enough to raise
the pericenter of all of the solids on typical dynamical time
scales. Although the central binary evacuates the region with $a
\approx$ 18--30~\RP\ in $\sim$ 1 yr, some tracers and their associated
solids are placed on lower $e$ orbits with $a \gtrsim$
30~\RP. Collisional damping continues to lower $e$ for these tracers;
they remain on fairly stable orbits on 10--100~yr times scales.  At
100--135~\RP, collisional damping is fast enough to slow the loss of
material compared to the model (b) and (c) calculations. On 100~yr
time scales, however, much of this material is still lost,

For systems with $q \approx q_a$, collisional damping enables
retention of a significant fraction of solids at $a \approx$
30--100~\RP. After one month (year), the system has 88\% (78\%) of its
initial mass. Although the loss of material reaches $\sim$ 35\% (38\%)
after 10 yr (100~yr), subsequent losses are small. Most of this
material has $a \approx$ 45--75~\RP, which overlaps the satellite zone
at 33--66~\RP.

\begin{figure}[t!]
\begin{center}
\includegraphics[width=5in]{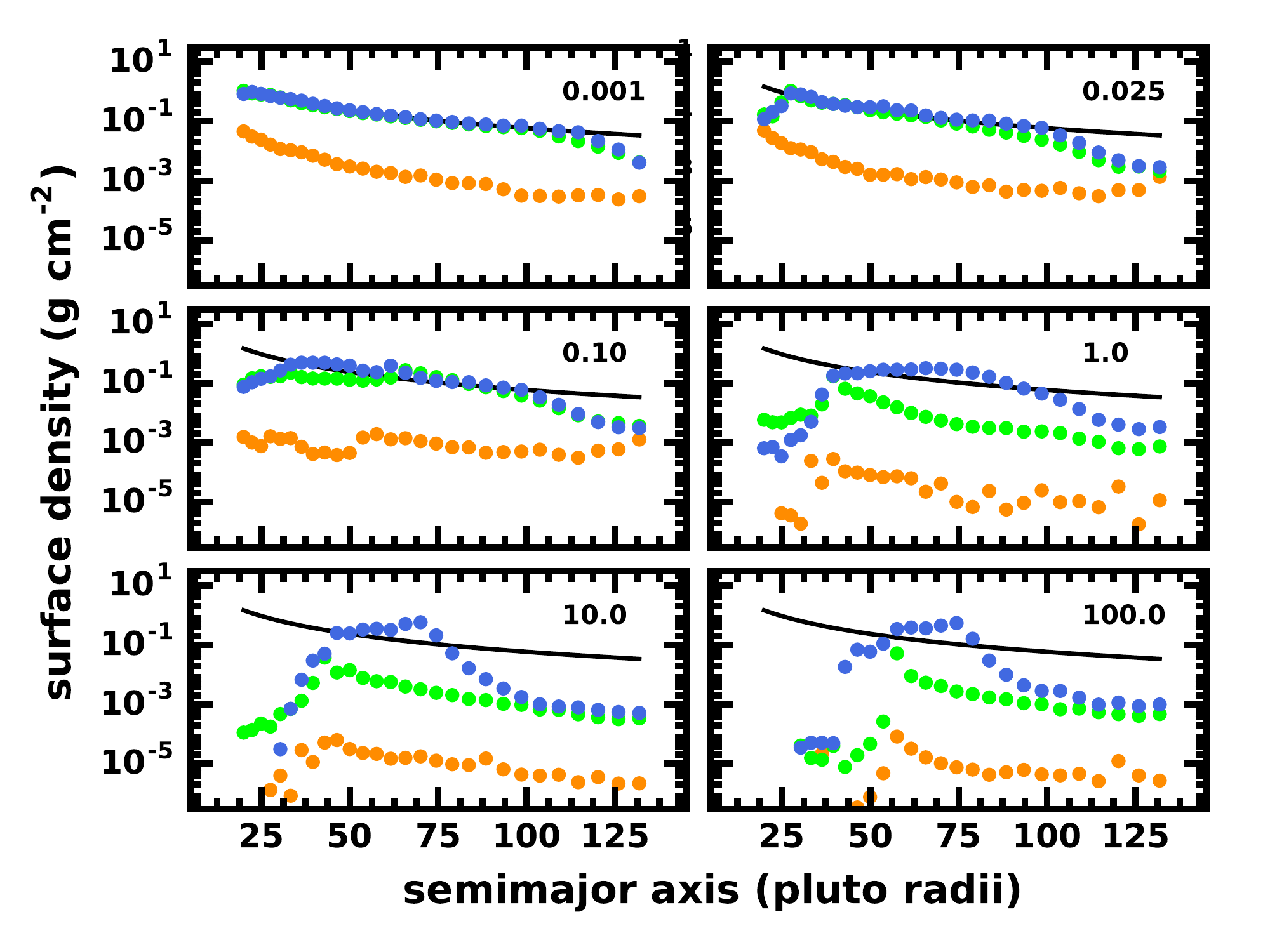}
\end{center}
\vskip -2ex
\caption{\label{fig: sigma1} Time evolution of the surface density in
  rings of solids at 19--135~\RP\ for systerms with $q = q_a$ (blue
  points), $q = q_b$ (green points), and $q = q_c$ (orange
  points). The solid black line in each panel indicates the initial
  surface density. The evolution time (in yr) is in the upper right
  corner of each panel.  Systems with $q \approx q_b - q_c$ lose
  nearly all of their solid material in 10--100~yr.  Systems with $q =
  q_a$ generate into a dense ring of solids at 50--80~\RP\ in
  10--100~yr; the cm- to m-sized solids then begin to grow into larger
  objects.  }
\end{figure}

Fig.~\ref{fig: sigma1} illustrates the time evolution of $\Sigma$ in
all three models.  In each panel, the thin black line plots the
initial surface density distribution, $\Sigma \propto a^{-2}$. The
points indicate the surface density in each annulus at the time (in
yr) indicated in the upper right corner of each panel for models with
$q = q_a$ (blue points), $q = q_b$ (green points), and $q = q_c$
(orange points).  The rapid evolution in the inner and outer disk is
apparent: it takes only 5--10 days to begin to remove solids at
20--30~\RP\ and at 110--135~\RP. In between these limits, collisional
damping tries to drive the solids to lower $e$ and lower $\imath$
before tracers are ejected or collide with Pluto or Charon. For the
model (b) and (c) parameters, collisional damping is too slow at
30--100~\RP. Although material is lost more slowly than solids in the
inner or outer disk, the reduction in surface density is steady at all
$a$.

In the $q = q_a$ calculation, the evolution concentrates solids in
annuli at 50--70~\RP. As with the other calculations, the central
binary steadily removes material in the inner disk and the outer
disk. The time scale for this evolution is 1--10~yr. At $t$ = 1 yr
(Fig.~\ref{fig: sigma1}, middle right panel), the surface density at
50--80~\RP\ is larger than the initial $\Sigma$. After another 9~yr,
the region of large $\Sigma$ is smaller; however, the region at
50--75~\RP\ still has a surface density somewhat larger the starting
point. After 100~yr (Fig.~\ref{fig: sigma1}, lower right panel), the
surface density at $\sim$ 55--80~\RP\ is at least as large as the
initial surface density.

When $q = q_a$ and $t$ = 30--200~yr, the surface density at
50--80~\RP\ oscillates slowly with time as tracers on modest $e$
orbits move into and out of the coagulation grid. Overall, the total
mass within the coagulation grid is fairly constant, varying by
roughly
a few percent from one time step to the next. Throughout this period,
the typical $e$ and $\imath$ of tracers gradually declines. At
100--200~yr, particles with $a \approx$ 50--80~\RP\ have typical $e
\approx$ 0.01 and $\imath \approx 10^{-4}$. This part of the swarm is
vertically thin and free from large dynamical interactions with
Pluto--Charon.

Once the solids settle down into low $e$, low $\imath$ orbits at
100--200~yr, mergers start to dominate collision outcomes. Typical
center-of-mass collision energies, $Q_c \approx 2 - 3 \times
10^4$~\ergg, are smaller than typical binding energies,
$\qdstar \approx 5 - 10 \times 10^5$~\ergg.
For computational simplicity, we do
not place tracers in mass bins with $r \gtrsim$ 1~m, preventing
\orch\ from following the growth of the largest objects. We therefore
terminate the calculation. Based on previous results \citep{kb2014pc},
we expect the swarm of 1~cm to 1~m objects to grow into a few 1~km
satellites in 100--1000~yr. Once satellites reach these sizes,
subsequent growth is chaotic; likely outcomes include 5--20~km
satellites with masses similar to the known moons in the satellite
zone.



{
To place the evolution of Fig.~\ref{fig: sigma1} in the
context of other giant impact calculations, we examine 
the evolution of $\Sigma$, $e$, and $\imath$ as functions 
of the `radius of the equivalent circular orbit' 
\citep[e.g.,][]{canup2004, canup2011, nakajima2014}
\begin{equation}
\label{eq: aeq}
a_{eq} = a (1 - e^2) ~ .
\end{equation}
For each tracer, the angular momentum is
$L^2 = G (m_P + m_C) a (1 ~ - ~ e^2)$; thus, 
$a_{eq}$ provides a way to track the angular momentum
evolution of the swarm.

At the start of each calculation, all of the solids have
a pericenter near Charon (the $q = q_a$ model), near Pluto
(the $q = q_c$ model), or in between (the $q = q_b$ model).
With $a_{eq} = q (1 - e)$ and $e > 0$, the initial $a_{eq}$ 
lies inside $q$. As the calculations proceed, collisional 
damping gradually reduces $e$ for many solids; in turn, 
$a_{eq}$ slowly increases. Throughout the evolution, the 
central binary ejects other solids from the system. Although
the $a_{eq}$ for this material also increases with time, 
these solids have $e \gtrsim$ 1 and a range of $\imath$. 
For short periods of time, unbound material contributes 
to the surface density for all $a_{eq} \gtrsim$ 0.

\begin{figure}[t!]
\begin{center}
\includegraphics[width=5in]{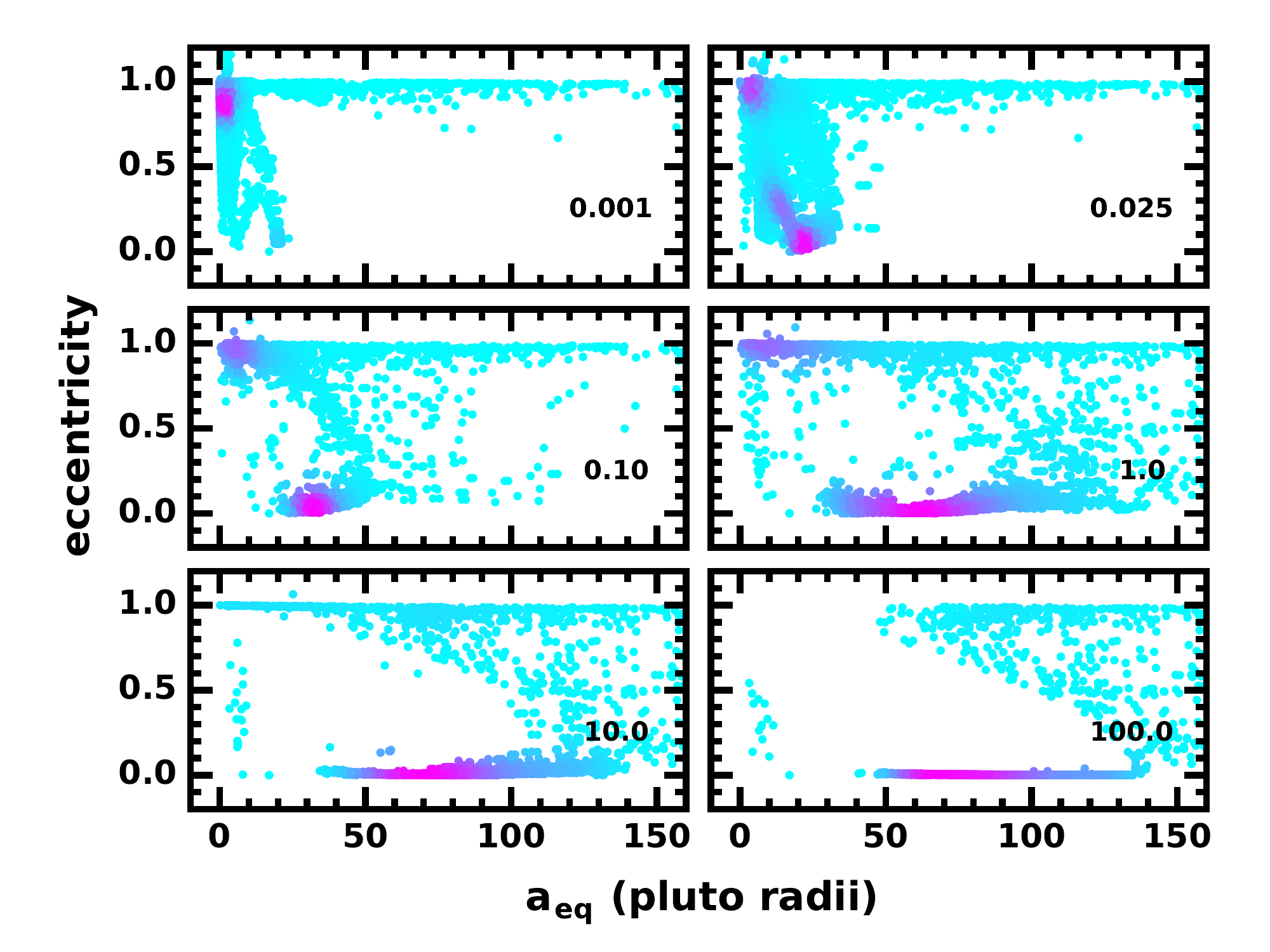}
\vskip -2ex
\caption{\label{fig: ecc1}
Time evolution of solids in the $(a_{eq}, e)$ plane for the
$q = q_c$ model. Cyan points indicate the positions of 
individual tracers. Magenta points denote regions of high 
density; the intensity of the color provides a measure of
the density. The evolution time in years is listed in the 
lower right corner of each panel.
}
\end{center}
\end{figure}

In the $q = q_c$ model, competition between collisional 
damping and dynamical ejections generates two distinct sets
of solids in the $(a_{eq}, e)$ plane (Fig.~\ref{fig: ecc1}).
At $t = 0$, all points lie on a nearly vertical line at
$a_{eq} \approx$ 0 between $e \approx$ 0.9 and $e \approx$ 0.99.
In a few hours, collisional damping creates the swath of points
with $a_{eq} \approx$ 5--20~\RP\ and $e \lesssim$ 0.9 
(Fig.~\ref{fig: ecc1}, upper left panel). At the same time, 
the central binary ejects solids along a locus with $e \approx$ 1
and $a_{eq} \approx$ 10--200~\RP. Although this rapid evolution 
is striking, the dynamical parameters of most tracers remain 
unchanged; these tracers lie in the magenta clump in the upper
left corner of this panel.

Over the next year, collisional damping drives a swarm of tracers
into the midplane of the central binary. After $\sim 10$ days,
a dense knot of tracers lies within a narrow ring at $a_{eq} \approx$ 
20~\RP\ (approximately the semimajor axis of Charon). 
Most of this material has $e \lesssim$ 0.1; some with larger $e$ 
approaches the ring from smaller $a_{eq}$ (Fig.~\ref{fig: ecc1}, 
upper right panel).  This ring expands to $a_{eq} \approx$ 30~\RP\ at 
$t$ = 0.1~yr (Fig.~\ref{fig: ecc1}, middle left panel) and then into 
a disk close to the satellite zone at $a_{eq}$ = 40--80~\RP\ after
a year of evolution (Fig.~\ref{fig: ecc1}, middle right panel). 

Throughout this period of disk formation, collisional damping drives
material from orbits with small $a_{eq}$ and large $e$ to those with 
larger $a_{eq}$ and smaller $e$. This evolution generates a slanted 
line of points in the Figure. At early times, this line appears to 
feed the dense ring at small $a_{eq}$. At later times, this line feeds 
the forming disk outside the dense ring. Eventually, there are no tracers 
in the region between the disk and the set of tracers on unbound orbits
(Fig.~\ref{fig: ecc1}, lower panels). Although we halted the evolution
after 100~yr, some of the tracers above the disk in the lower panels
will join the disk at large $a_{eq}$. 

As many tracers join the disk, the central binary ejects others. These
tracers form a line with $e$ = 1 and all $a_{eq} \gtrsim$ 0. The actions
of collisional damping and dynamical ejections gradually deplete the knot
of tracers at $(a_{eq}, e)$ = (0, 1) in Fig.~\ref{fig: ecc1}. After
$t \approx$ 10~yr (100~yr), there are few (no) tracers in this knot. 
Another few hundred years of evolution will probably eliminate all tracers
with $e \approx$ 1 from the diagram.

\begin{figure}[t!]
\begin{center}
  \includegraphics[width=5in]{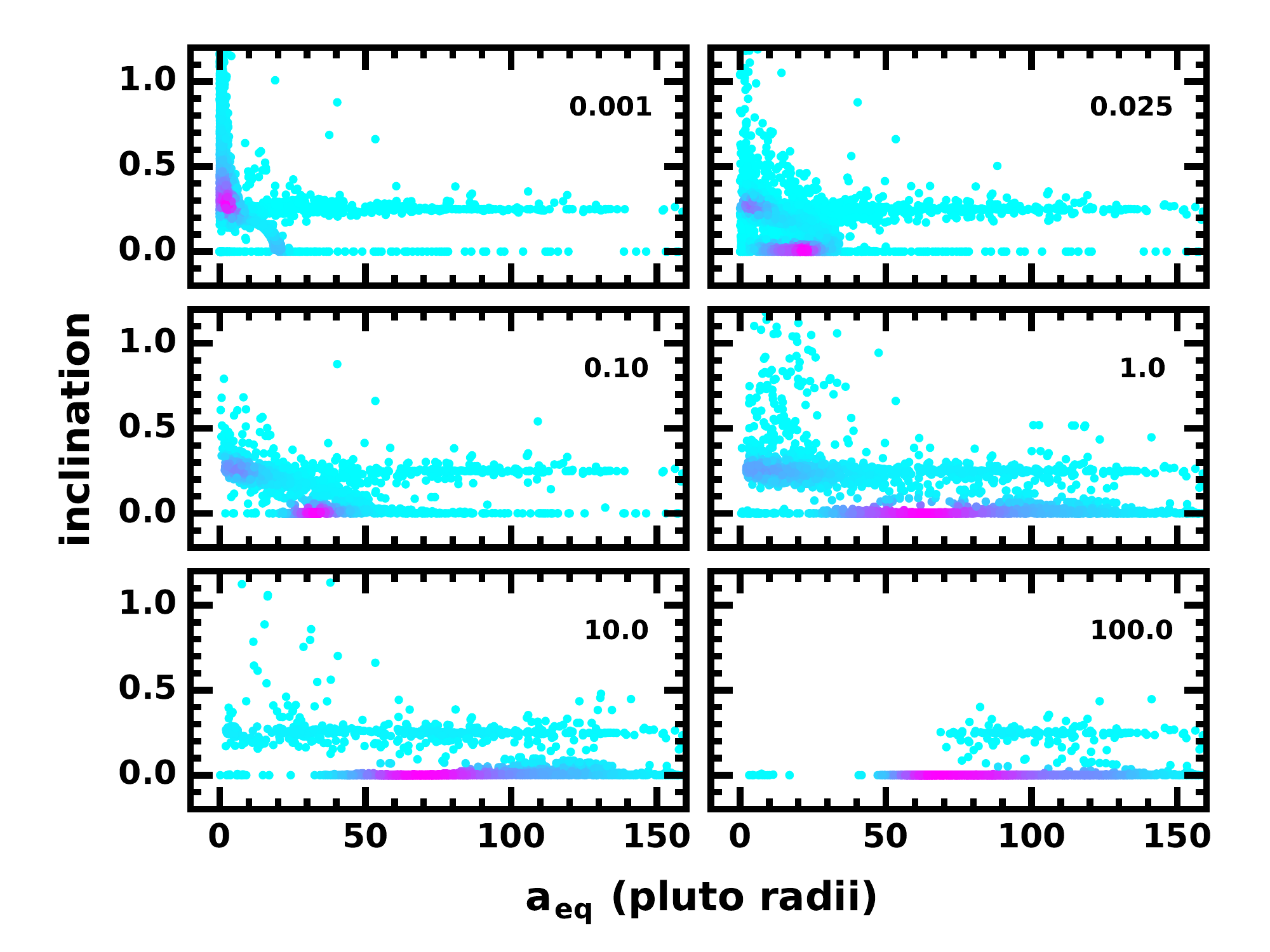}
  \end{center}
\vskip -2ex
\caption{\label{fig: inc1}
As in Fig.~\ref{fig: ecc1} for the inclination.
}
\end{figure}

Fig.~\ref{fig: inc1} repeats Fig.~\ref{fig: ecc1} for the inclination
of particles in the $q = q_c$ model. At $t$ = 0, all tracers have 
$a_{eq} \approx$ 0 and $\imath \approx$ 0.25. Within a few hours,
ejections from the central binary establish a set of tracers with small 
$a_{eq}$ and large $\imath$ (Fig.~\ref{fig: inc1}, upper left panel).
Another set of ejected particles has $a_{eq} > 0$ and $\imath \approx$ 0.25.
A third group has damped into the orbital plane ($\imath \approx$ 0). As 
with $e$, there is a set of tracers along a line from the dense knot at
$(a_{eq}, \imath)$ = (0., 0.25) to the midplane. 

As the evolution proceeds, tracers continue to be ejected along lines with
$a_{eq} \approx$ 0 or $\imath \approx$ 0.25. Several tracers not shown in
the Figure are ejected on retrograde orbits relative to the central binary.
From $t$ = 0.001~yr to $t$ = 1~yr, more and more tracers are ejected along
the line with $\imath \approx$ 0.25 compared to the line with $a_{eq} \approx$ 0.
At $t$ = 10--100~yr, tracers remaining to be ejected have $a_{eq} \gtrsim$
50--100~\RP\ and $\imath \approx$ 0.25. After another few hundred yr, the
central binary will eject these tracers.

As most tracers disappear from the system, collisional damping drives a 
second set of tracers into the midplane. At early times (Fig.~\ref{fig: inc1},
upper right panel), tracers from the initial knot feed into the growing ring
at $a_{eq} \approx$ 20~\RP. Expansion of this ring into a disk accompanies
the depletion of the initial dense knot of tracers (Fig.~\ref{fig: inc1},
middle panels). After most tracers have been ejected, a dense disk of tracers
remains in the midplane of the binary (Fig.~\ref{fig: inc1}, lower panels).
Much of this disk lies within the satellite zone.

The evolution of tracers in the $q = q_b$ model follows the evolution in 
Figs.~\ref{fig: ecc1}--\ref{fig: inc1}. Having somewhat smaller (larger) 
$e$ ($q$), these tracers initially form a dense knot in $(a_{eq}, e)$ 
space at somewhat larger (smaller) $a_{eq}$ ($e$). Collisional damping and
dynamical ejections then transform this knot into a dense disk of solids 
in the midplane of the binary and a swarm of material flowing out of the
system. Although some material is ejected at large $\imath$ with respect 
to the midplane, most flows out through a disk-shaped volume with its
equator in the binary midplane and an opening angle of $\imath \approx$ 0.25.
Compared to the $q = q_c$ model, more material remains in the disk and
ejections are less frequent.

\begin{figure}[t!]
\begin{center}
  \includegraphics[width=5in]{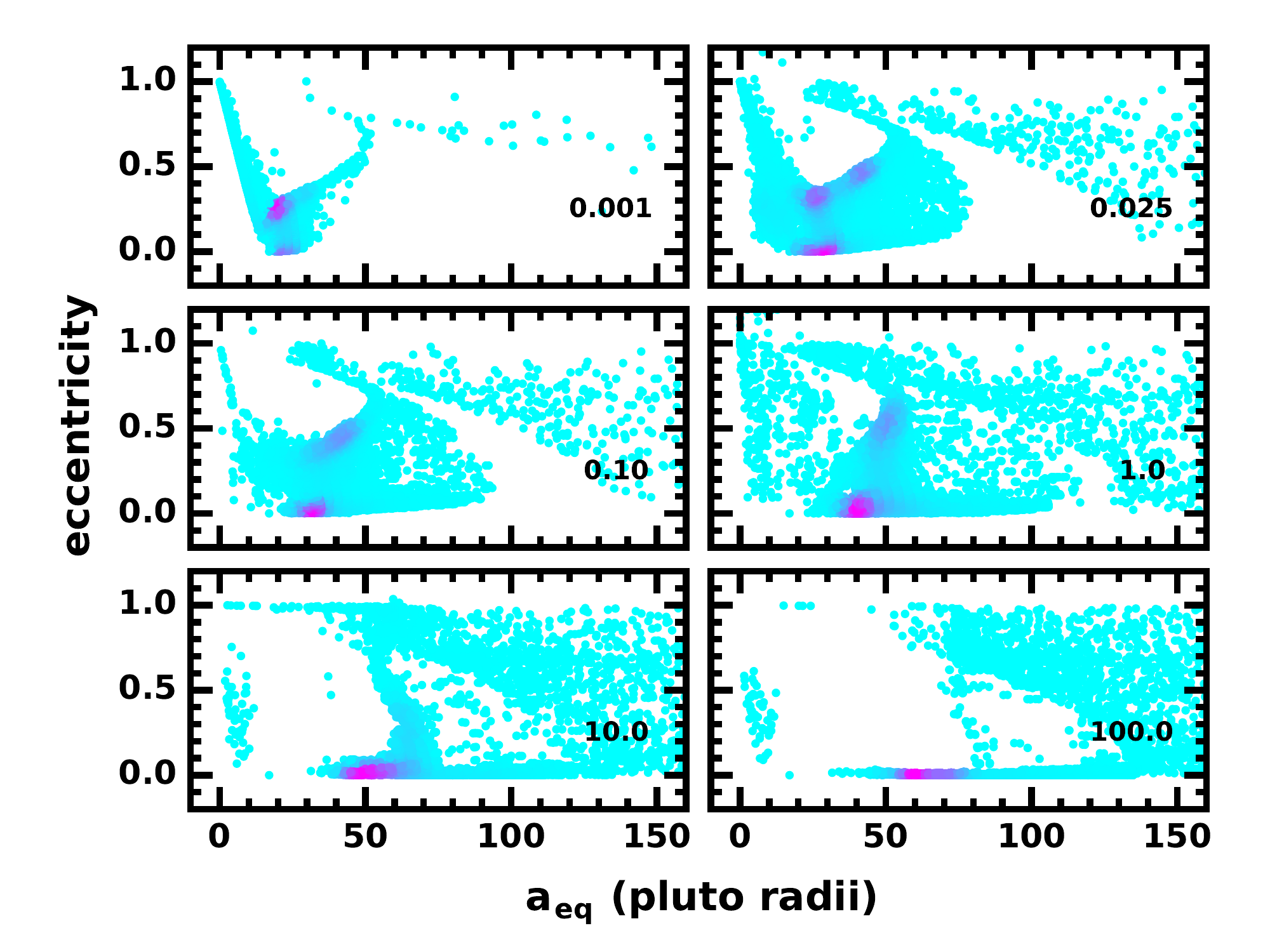}
  \end{center}
\vskip -2ex
\caption{\label{fig: ecc2}
As in Fig.~\ref{fig: ecc1} for the $q = q_a$ model. }
\end{figure}

When $q = q_a$, the evolution of tracer parameters as a function of
$a_{eq}$ changes dramatically (Fig.~\ref{fig: ecc2}). In this model,
material initially has a broad range of eccentricity, $e \approx$
0.25--0.9. In the $(a_{eq}, e)$ plane, tracers occupy a line extending
from $(a_{eq}, e)$ = (20, 0.25) to $(a_{eq}, e) \approx$ (0,
0.9). Within a few hours, damping places some tracers within a narrow
ring at $a_{eq} \approx$ 20~\RP\ in the midplane of the binary.  The
central binary ejects others into a swath with $a_{eq} \approx$
20--50~\RP\ and $e \approx$ 0.3--0.7 (Fig.~\ref{fig: ecc2}, upper left
panel).

At later times, damping and dynamical ejections continue to evolve
tracers along different paths. After 0.025 yr, there is a dense
concentration of solids in a broad ring with $a_{eq} \approx$
20--30~\RP\ (Fig.~\ref{fig: ecc2}, upper right panel).  Damping also
translates an ensemble of tracers from their initial $(a_{eq}, e)
\approx$ (20, 0.25) to (25, 0.25). In contrast, interactions with the
central binary displace another set of tracers towards larger $a_{eq}$
and larger $e$: there is a dense clump of tracers at $(a_{eq}, e)$ =
(40, 0.5) in the midst of a large group on its way out of the system.

During the next 0.1--1~yr, these two groups of solids diverge more and more 
dramatically in the $(a_{eq}, e)$ plane (Fig.~\ref{fig: ecc2}, middle panels).
The dense clump of tracers at $(a_{eq}, e)$ = (20, 0.25) vanishes. The dense 
clump in the midplane grows in mass and expands to $a_{eq} \approx$ 40~\RP.
Although several tracers remain behind at small $a_{eq}$, most of the remaining
tracers move to larger $a_{eq}$ at ever-increasing $e$.

By 10--100~yr, disk formation within the satellite zone is nearly
complete (Fig.~\ref{fig: ecc2}, lower panels). Although the disk
extends from 30~\RP\ to beyond 150~\RP, the densest concentration of
tracers at 55--75~\RP\ overlaps the satellite zone. Aside from several
dozen tracers at small $a_{eq}$ and small $e$, few tracers have orbits
with $a_{eq} \lesssim$ 50--70~\RP\ and $e \gtrsim$ 0.05.  At larger
$a_{eq}$, tracers are either within the disk, damping into the disk at
$a_{eq} \gtrsim$ 150~\RP, or leaving the system on high $e$ orbits.

\begin{figure}[t!]
\begin{center}
  \includegraphics[width=5in]{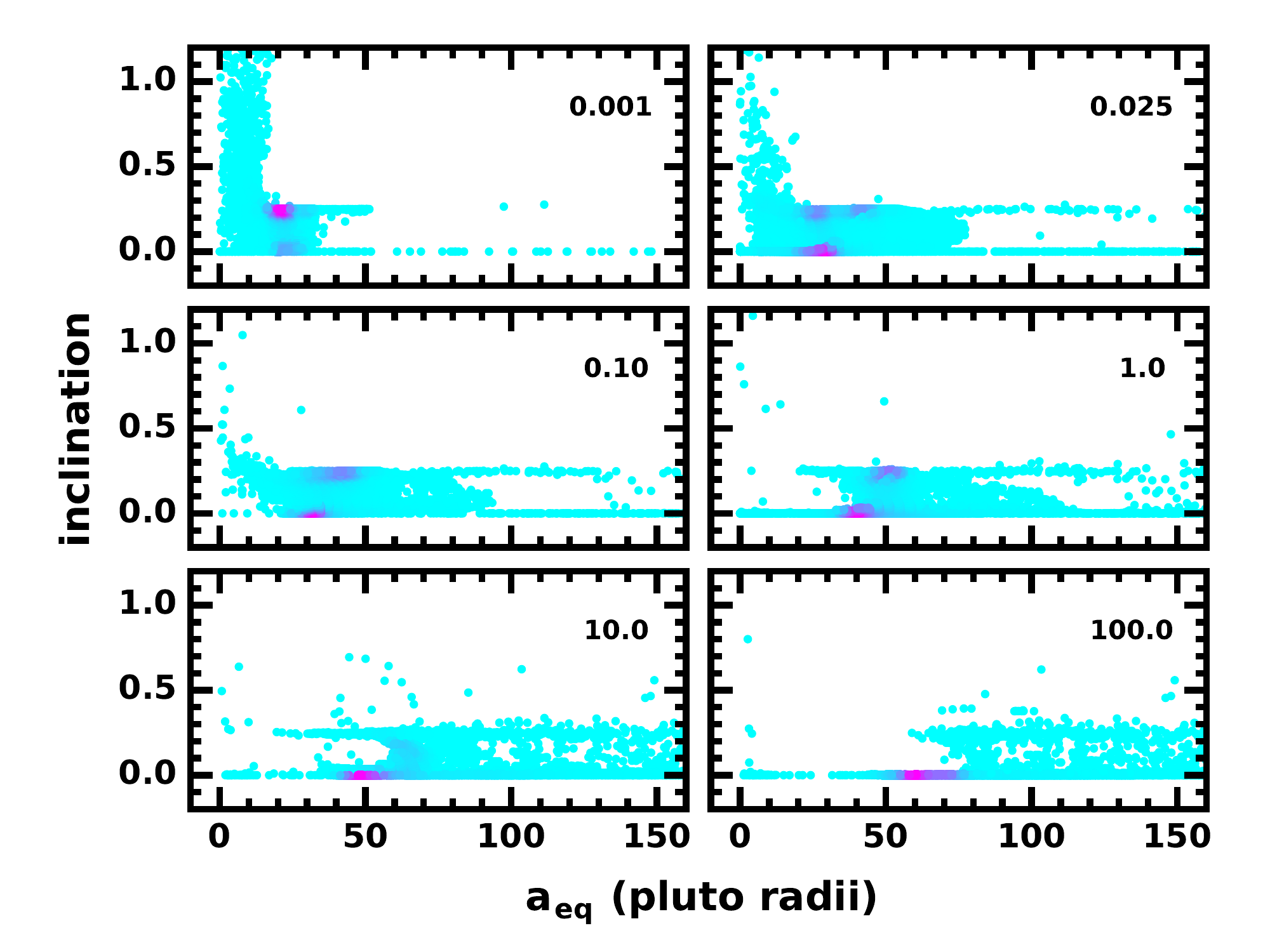}
  \end{center}
\vskip -2ex
\caption{\label{fig: inc2}
As in Fig.~\ref{fig: ecc2} for the inclination.
}
\end{figure}

Fig.~\ref{fig: inc2} illustrates the time evolution of tracers in the
$(a_{eq}, \imath)$ plane when $q = q_a$. Starting from a configuration
where solids lie in a horizontal line with $a_{eq}$ = 0--20 and 
$\imath$ = 0.25, the central binary ejects a few tracers along retrograde
orbits and pushes a much larger subset of tracers into high inclination 
orbits (Fig.~\ref{fig: inc2}, upper left panel).

Collisional damping drives another large group into a low density ring 
in the orbital plane of the central binary. After only 9--10 days
(Fig.~\ref{fig: inc2}, upper right panel), many tracers have low 
inclination orbits ($\imath \approx$ 0) extending from $a_{eq} \sim$ 0
to $a_{eq} \gtrsim$ 150~\RP. The density of tracers in the narrow ring
at $a_{eq} \approx$ 20--35~\RP\ is then larger than the density in the
original knot at $(a_{eq}, \imath)$ = (20, 0.25) or the ensemble of
tracers ejected on high eccentricity orbits with large $a_{eq}$ and 
$\imath \approx$ 0.25.

As the calculation proceeds, tracers separate into two distinct groups.
In the orbital plane of the binary, the position of the narrow ring 
moves from $a_{eq} \approx$ 20~\RP\ to $a_{eq} \approx$ 40~\RP\ in 1~yr
(Fig.~\ref{fig: inc2}, middle panels). At later times, the ring expands
into a disk which extends from 30~\RP\ to $\gtrsim$ 100~\RP. The densest
part of this ring lies within the satellite zone (Fig.~\ref{fig: ecc2},
lower panels).

As the disk forms, the central binary ejects tracers in all directions.
Although some tracers pass out of the system on high inclination orbits,
most follow lower inclination trajectories with $\imath \lesssim$ 0.25.
During the first 0.1--1~yr, the binary evacuates inner regions with
$a \lesssim$ 30~\RP. Tracers at larger $a$ on longer period orbits 
encounter the binary less frequently and are ejected on longer times scales
of 10--100~yr. After 100~yr, there are only a few tracers with 
$a_{eq} \lesssim$ 70~\RP. High inclination tracers at larger $a_{eq}$ 
will be ejected over the next few hundred years.

The evolution outlined in Figs.~\ref{fig: ecc1}--\ref{fig: inc2}
illustrates the transformation of a high eccentricity, high
inclination swarm of solids into a vertically thin disk that overlaps
the satellite zone. To show how the surface density evolves, we assign
tracers to discrete bins in $a_{eq}$, sum the mass carried by each
tracer, and divide by the area for each bin.  Fig.~\ref{fig: sigma2}
shows snapshots of the surface density distribution $\Sigma(a_{eq})$
for the same epochs shown in Fig.~\ref{fig: sigma1}. At the start of
each sequence, $a_{eq} \approx q$; all of the solids are bunched up
close to Charon.  In the $q = q_a$ model (Fig.~\ref{fig: sigma2}, blue
points), 0.001~yr of collisional damping generates a ring with high
surface density at 20--50~\RP\ and an extended region with low surface
density at 50--150~\RP. Over 10--100~yr, the dense ring expands
outward, reaching 45--75~\RP. Inside this ring, the surface density
drops by a factor of 100--1000. Outside, $\Sigma$ grows by a factor of
100.

Although collisional damping tries to generate a dense ring in the
other models, dynamical ejections are more efficient. In the $q = q_c$
model (Fig.~\ref{fig: sigma2}, orange points), some material lies
within the low $\Sigma$ extended region. Over time, the surface
density in this model gradually declines. In the $q = q_b$ model, a
small dense ring is well-defined at 0.001--0.1~yr. However, the
surface density slowly declines with time, leaving behind an extended
disk with low surface density at 50--150~\RP.

\begin{figure}[t!]
\begin{center}
  \includegraphics[width=5in]{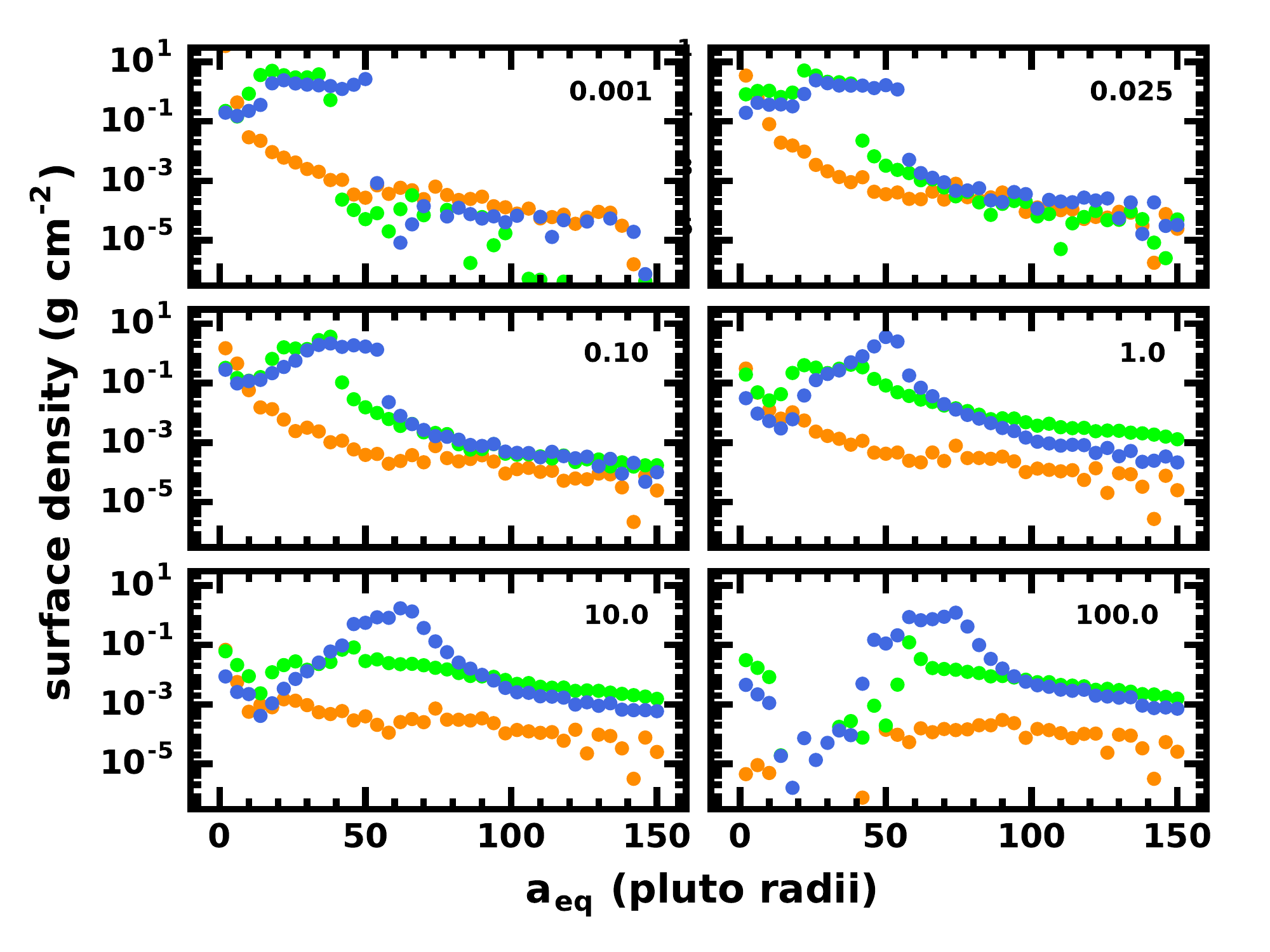}
  \end{center}
\vskip -2ex
\caption{\label{fig: sigma2}
Snapshots of the surface density in calculations with
$q = q_a$ (blue points), $q = q_b$ (green points), and 
$q = q_c$ (orange points) at times (in yr) indicated in
the upper right corner of each panel. After 100~yr,
the $q = q_a$ model has a dense ring at 50--80~\RP\ within
a lower density disk at 40--150~\RP. This ring is stable.
In the other models, the surface density within the ring and
the disk gradually decline with time.
}
\end{figure}

} 

\subsection{Outcomes}

Outside the dense ring of solids at 50--80~\RP, the evolution has
several likely outcomes.  Inside 35--40~\RP, where the surface density
is low, the eccentricity is small $e \approx$ 0.01--0.02. With low $e$
and little mass, these solids will either remain in their present
state or collide with material at somewhat larger distances. Either
way, satellite formation at 30--40~\RP\ is unlikely. Beyond 
80--90~\RP, roughly half of the mass is in small (large) solids with
$r \lesssim$ 3--5~cm ($r \gtrsim$ 5--10~cm) and $e \lesssim$ 0.01 ($e
\gtrsim$ 0.1).  With such a small surface density, collisions are
rare; damping is slow.  Although safe from interactions with Charon,
these solids are unlikely to merge into larger objects but may merge
with solids in the dense ring. Collectively, the solids at
30--40~\RP\ and at 90--130~\RP\ have a total mass comparable to the
mass of Styx, $m \approx 7 \times 10^{17}$~g. With such little mass,
this material will have little impact on the formation of satellites
in the dense ring.

Once growth is complete at 50--80~\RP, the low density material at
30--40~\RP\ and at 90--130~\RP\ may play a role in shaping the final
configuration of the satellites.  With more material in the outer disk
than the inner disk, satellites may migrate inward toward the central
binary \citep{kb2014pc,bk2015pc}. Satellites cannot migrate into the
unstable region inside 30~\RP\ due to the lack of material. However,
migration could place small satellites -- like Styx -- inside the
dense ring at 50--80~\RP.

Several test calculations suggest the formation of a dense ring in the
satellite zone is the inevitable outcome of evolution when material
has $q \approx a_c$. Systems with steeper (shallower) surface density
gradients allow ring formation somewhat closer (farther) away from the
central binary than the ring in Fig.~\ref{fig: sigma1}. In a future
study, we plan to consider how outcomes depend on the initial
conditions in more detail.

\subsection{Summary}

Calculations with \orch\ demonstrate that the evolution of ejecta from
an impact with Pluto or Charon is a competition between collisional
damping and removal by the central binary. Unless impact debris is
launched from Charon's orbital distance, less than a few percent of
the debris remains after $\sim$ 100~yr.  Satellite formation is then
unlikely unless the original ejecta mass is significantly larger than 
$\Msats \sim 10^{20}$~g.  A launch from Charon is much more promising.
After $\sim$ 100 yr of collisional evolution, more than 60\% of the 
debris remains in low eccentricity orbits around the central binary.
This material is constrained to a region that is representative of the 
satellite zone, with little mass outside of it. The resulting satellite 
system \citep{kb2014pc} will be truncated, consistent with the absence
of moons beyond Hydra.

Natural extensions of these calculations include a consideration of
different particle densities, initial size distributions, and orbital
configurations, as suggested by the analytical estimates in
\S\ref{sec:formdisk}.

\section{Discussion}\label{sec:discuss}

The goal of this work is to explore models for creating a reservoir of
mass for the small satellites of Pluto and Charon from a Charon-TNO
impact.  Above (\S\S\ref{sec:formdisk} and \ref{sec:orch}), we
consider mechanisms and conditions for settling ejecta from a direct
hit on Charon into the satellite zone around the binary.  In this
section, we discuss these assessments in the broader context of the
TNO impact. Our main concerns include how big the impactor must be,
given the efficiency of the delivery of impact ejecta to the satellite
zone, and the likelihood of such a collision with Charon over the
history of the solar system.
 
As we describe in \S\ref{sec:formdisk}, impact debris must orbitally
damp quickly to settle into the satellite zone before being ejected or
accreted by the binary.  The efficiency of this process is key to
understanding the impact event itself. When damping is driven by
collisions (\S\ref{sec:collide}), dynamical cooling is rapid but only
if debris particles are small and numerous enough to support a high
collision rate (cf.~Eq.~(\ref{eq:tcol})). When the TNO impact launches
just enough mass into the satellite zone to build the satellites, then
the debris particles must be less than about 10 meters in radius to
settle there through collisional processes. When the debris particles
are larger, the ejecta must have a greater total mass for settling to
occur.  We explore the connection between ejecta and impactor in more
detail in \S\ref{sec:massimpactor}.

Gas drag within a cloud of vaporized ice produced in the original
impact with Charon offers another possible pathway for delivering debris
to the satellite zone (\S\ref{sec:gas}). Small debris particles can
become entrained in the gas and settle to the midplane as the cloud
evaporates. A cloud with a mass of $0.1\Msats$ can entrain
sub-millimeter grains; trapping larger grains requires more gas. If
the debris initially consists of particles that are too large to be
entrained, a collisional cascade may grind the debris into particles
of the right size.  Depending on how the particle sizes evolve, and on
the uncertain details of the gas dynamics, a majority of the small
debris may be captured in the gas cloud. This effect could help reduce
the amount of ejecta needed to get mass to the satellite zone.

In both scenarios, an impact on Charon generates a reservoir of small
particles in a thin, dynamically cold circumbinary disk.  These solids
are capable of trapping or eroding larger debris fragments. In models
where the mass of small debris particles is comparable to the total
mass of the Pluto-Charon satellites, the maximum size of particles
that can be trapped is about 10~meters.  Depending on the size
distribution and orbital elements of these larger bodies, trapping
will further increase the mass available to build Kerberos and the others.

With or without these larger ``seed'' fragments, coagulation processes
lead to the growth of the satellites \citep[e.g.][]{kb2014pc}.  This
outcome is possible because the debris around Pluto-Charon is well
outside of the Roche limits of the central bodies. In contrast, the
small particles that make up Saturn's rings lie within the planet's
Roche limit, where tidal forces prevent them from sticking
together. The rings are long-lived for this reason. We conclude that a
circumbinary disk produced by a Charon-TNO impact will inevitably form
satellites.

\subsection{The mass of the impactor}\label{sec:massimpactor}

To make a Charon-impact scenario plausible, the giant cratering event
from the TNO collision must eject enough material to account for the
satellites.  We estimate the necessary mass with guidance from the
simulations in \S\ref{sec:formdisk}.  The tracer particles in these
simulations represent the high-speed impact ejecta capable of escaping
Charon's Hill sphere. Only about 1\%\ of these particles end up in the
satellite zone after damping ``in-place'' around their osculation
semimajor axis. This result suggests that the mass of the high-speed
ejecta may need to be as high as $\sim 100 \times \Msats$.  However,
because the number density of particles is high in the satellite
zone, bound debris that passes through it can get trapped there
(\S\ref{subsec:trap}).  Then, as many as 10\%\ of the tracers may end
up in the satellite zone (Fig.~\ref{fig:tsurvive}). In this case, the
minimum mass of the high-speed ejected debris is about $10^{21}$~g, or
$10 \times \Msats$.

The mass of the impactor needed to eject {$\sim$}${10}^{21}$~g depends
on uncertain characteristics of the impactor's composition and
structure, along with the physics of the collision, including the
impactor's speed and angle of impact. To address the impact geometry,
we consider limiting cases: In a direct hit, where the projectile
impacts Charon head-on, the impactor is obliterated, launching debris
and leaving a giant crater. In a surface-skimming event, where the TNO
plows into Charon at an oblique angle, material is ejected along the
impactor's general direction of travel. We assume that the impactor is
also destroyed in this case as well, although debris may be launched
similarly even if the impactor survives and continues on its way after
it scrapes debris from the surface of Charon \citep[as in the
  ``hit-and-run'' scenario for the formation of the Pluto-Charon
  binary itself;][]{canup2005}.

The direct-hit scenario is likely inefficient, requiring an impactor
with substantially more mass than is currently in the satellites. The
oblique case may be efficient, involving a lower-mass TNO, if the
impact hits Charon from a direction that puts debris on low
inclination orbits in the satellite zone. The chance that the impact
geometry (impact parameter, direction of travel relative to Charon's
orbit around Pluto) is just right may be low compared with the
geometry required of a direct impact, for which the simulations suggest that
the location of the impact on the surface of Charon is not a big
factor. However, because smaller TNOs are significantly more common
than larger ones (see below), the surface-skimming impact scenario may
be at least as likely as a direct hit by a larger TNO.

In this preliminary work we focus on the direct-hit scenario. Our
simulations are idealizations of this case, and elegant scaling laws 
provide quantitative predictions for aspects of cratering events
\citep[e.g.,][]{hols1994}. One of these laws yields the impactor mass
required to eject debris at the speeds needed to populate the
satellite zone:
\begin{equation}\label{eq:mscale}
\frac{M(v)}{\Mi}  = 
     {C} \left(\frac{v}{\vi\cos\thetai}\right)^{-3\mu}
     \left(\frac{\rhoi}{\rhoc}\right)^{1-3\nu}
\end{equation}
where $M(v)$ is the mass ejected with speed greater than $v$
\citep{hols1994, housen2011}.  The parameters $\Mi$, $\vi$, and
$\thetai$ are the impactor's incoming mass, speed, and incidence
angle, respectively. The constant $C$ and index $\mu$ depend on
properties of the target material; the index $\nu$ accounts for
the effect of any difference in mass density between target ($\rhoc$)
and impactor ($\rhoi$). We caution that the power-law dependence is
strictly valid for the case of a small impactor and a planar target.
In our case, the mass ratio of projectile and target is expected to be
small ($<$1\%), but the ratio of radii ($\sim$10\%) is not.

To apply Equation~(\ref{eq:mscale}), we assume that the target,
Charon, is ``well-baked'' by tidal interactions with Pluto and thus
has low porosity. Then, the index $\mu = 2/3$, consistent with the
velocity distribution in our simulations if tracers represent
identical masses \citep[cf.][]{housen2011, shuvalov2009, svet2011}.
While we usually treat the density at Charon's surface to be the same
as that of the impactor, we set the index $\nu = 0.4$
\citep{housen2011}. Our choice for the constant $C$ is 0.03,
consistent with hydrodynamical simulations of other non-porous
material \citep[e.g.,][]{svet2011}. We assume an angle of incidence
$\thetai = 0^\circ$ (a head-on collision), and an impact speed of $\vi
= 2$~km/s, typical of the relative speed between Plutinos and other
TNOs \citep{delloro2013}.  Finally, we set the threshold speed $v$ to
be 95\% of the surface escape velocity of Charon, as in our tracer
simulations.

Putting these values together, the impactor mass is
\begin{eqnarray}\nonumber
\Mi & \approx & \frac{M(v)}{C}
  \left(\frac{v}{\vi\cos\thetai}\right)^{3\mu}
     \left(\frac{\rhoi}{\rhoc}\right)^{3\nu-1}
     \\
     \label{eq:Miest}
  \ & \approx &  3.1 \times 10^{21}
 \MvtenMsats \left[\frac{C}{0.02}\right]^{\!-1}
 \vitwokms^{2} \left[\cos \thetai\right]^{2}
 \left[\frac{\rhoi}{\rhoc}\right]^{0.2} \ \text{g}
\end{eqnarray}
which corresponds to an impactor radius of 100~km for a density of
$\rhoi = 1$~g/cm$^3$.  The efficiency of ejecta production increases
with impactor density and collision speed. For example, raising $\vi$
to 3~km/s cuts the impactor mass in Equation~(\ref{eq:Miest}) by more
than a factor of two. An increase of closing speed to 4~km/s would
allow a 60~km impactor to deliver $10 \times \Msats$ of mass to the
satellite zone.  Conversely, increasing the incidence angle reduces
the efficiency. 

Our expectation is that an impact between Charon and an object with
mass derived from Equation~(\ref{eq:Miest}) would produce considerably
more debris than the point-mass scaling formalism suggests
\citep[see][]{svet2011, arakawa2019}. One reason is that ejecta is
launched from a convex target, not a planar one. It is plausible that
a 100-kilometer impactor could generate at least its own mass in
high-speed ejecta. By the same token, the parameters adopted in other
work lead to predictions with much less debris for the same impactor mass
\citep[e.g.,][]{bierhaus2015}.

If we have overestimated the debris production of a 100-kilometer
impactor, and if only 1\%\ of the high-speed ejecta settles in the
satellite zone, then the impactor must be disturbingly large, over
200~km in radius. With a mass well over $10^{22}$~g, this object
could significantly deflect Charon in its orbit around Pluto, possibly
initiating a second phase of tidal evolution. For this reason, and
because such a high-mass TNO is comparatively rare in the outer solar
system, a low-mass impactor plays a more credible role in the
scenarios hypothesized here.

As an alternative to a direct hit by a 100-kilometer TNO, an oblique,
surface-skimming impact from a smaller body might also produce enough
debris to make the satellites. For example, a 30-km impactor has a 
disruption threshold $\qdstar$ of about $5\times 10^8$~erg/g \citep[Fig.~5
  therein]{benz1999}; its specific kinetic energy is $2\times 10^{10}$ ergs/g. 
A surface-skimming impact (an oblique impact with ``erosion'' in 
\citealt{lein2012}) has the potential to completely break up the impactor 
and impart enough kinetic energy to launch about ten times its mass from 
Charon's crust along its direction of travel.  In an event where the ejecta 
are ``beamed,'' the impact geometry matters. A bad strike, with an impactor 
that hits Charon on a path aimed at Pluto, could beam all the debris right 
into the larger planet. A fortunate strike has a trajectory in the plane of
the binary, tangential to Charon's orbit, and intersecting with Charon at a 
point opposite from Pluto. Then debris would then be on prograde, low 
inclination orbits.

In summary, the production of the Pluto-Charon moons from a direct hit
probably requires an impactor that is 10--100 times the mass of the 
satellites, or 50-100~km in radius, depending on the impact speed. Higher 
speed, near the tail of the TNO velocity distribution, allows the smaller
mass. An even smaller TNO (a radius of 30~km or even less) may be sufficient 
in a fortuitously aimed surface-skimming event.

\subsection{The size distribution of the ejecta}\label{sec:size}

The success of a TNO impact as the origin of the Pluto-Charon
satellites depends on having most of the impact ejecta in the form of
small particles. Particles with radii less than a centimeter damp 
rapidly. Larger objects with radii of 1--10~m and total mass of 
0.3--1.0~$\Msats$ initiate a collisional cascade to produce cm-sized 
objects, which then damp rapidly.  In a direct-hit scenario, small 
sizes for the debris are plausible. From studies of impact fragmentation, 
nearly all of the mass of debris is in small objects
\citep[e.g.,][]{melosh:dynamicfrag}; meter-size bodies may even be
the most prevalent \citep{melosh1984}.  For a surface-skimming impact,
the specific kinetic energy exceeds the disruption threshold by two
orders of magnitude; as in \citet[for example]{benz1999}, we do not
expect large fragments from the projectile to survive.

\subsection{Likelihood of a Charon-impact event}

A key factor for the viability of our model is the probability that a
large TNO has collided with Charon since the formation of the binary.
Toward estimating this likelihood, we define $\Nfifty$ as the total
number of objects with radius $r > 50$~km in the Kuiper belt
today. From the observed brightness distribution of TNOs, \citet[see
  Table 5 therein]{petit2011} estimate $\Nfifty \approx 10^5$. Since
the Kuiper belt was more massive at early times \citep[e.g.,][and
  references therein]{kenyon2002}, we introduce a scale factor, $\fkb
\geq 1$, giving the ratio of mass in the Kuiper in some past epoch
relative to its mass today, roughly $0.01 M_\oplus$
\citep[e.g.,][]{fraser2014, pitjeva2018}. When the solar system was
young, the mass was as high as $1$--$10 M_\oplus$
\citep[e.g.,][]{kl1999b, kenyon2002, levison2008, booth2009,
  schlicht2011, schlicht2013, kb2014pc}, implying $\fkb \approx$
100--1000. Collision probabilities for specific solar system bodies
scale the same way.  Large impacts on Charon (and Pluto) were as much
as 100--1000 times more frequent 3--4 Gyr ago than those today.

The rate of collisions between Charon and these large TNOs depends on
several factors in addition to their total number, $\fkb \Nfifty$. The collision
cross sections of Charon and the impactors play a part, along with the
orbital characteristics that determine how frequently orbits
cross. Formally the collision rate is
\begin{equation}\label{eq:pcollide}
  \eta \approx N(r,t) P_i \pi (r + \RC)^2 \approx 0.03
  \fkb \left[\frac{\Nfifty}{10^5}\right]
  \left[\frac{\ri}{{50\ \text{km}}}\right]^{1-q} \left[\frac{P_i}{2
      \times 10^{-22}\ \text{km$^2$/yr}}\right]
  \left[\frac{\ri+\RC}{650\ \text{km}}\right] \ \text{Gyr}^{-1}
\end{equation}
where $N(r,t) \sim r^{1-q}$ is the cumulative size distribution for
radius $r$ and time $t$, $q$ is 3.5--8 as derived from the brightness
distribution and dependent on the specific TNO (sub)population
\citep[e.g.,][]{petit2011, gladman2012, shankman2013, fraser2014,
  adams2014, schwamb2014, alexandersen2016, shankman2016, lawler2018},
and $P_i$ is the mean intrinsic probability of collision
\citep{wetherill1967, greenberg1982}, which comes from assessments of
the distribution of orbit elements of the target and the potential
impactors. Our numerical choice for $P_i$ is for collisions between
Plutinos and all TNOs \citep{delloro2013}.

We conclude from Equation~(\ref{eq:pcollide}) that a collision between
Charon and an impactor with $r \gtrsim 50$~km is possible over the
last few Gyr, with a likelihood at the 10\% level. An impact event
with a more substantial body, with $r \gtrsim 100$~km, is much less
likely. Even with a generously shallow size-distribution slope with
$q=4$, a collision can be ruled out at the 98\%\ level.

Based on these event frequency estimates, a direct hit between Charon and 
a TNO with a radius $\ri$ = 50~km is a plausible origin for the building
blocks of the small satellites. However, the model requires a large amount 
of impact debris compared with the predictions of the cratering scaling law,
Equation~(\ref{eq:Miest}). Greater efficiency in the production of high-speed
ejecta remains a possibility, given the uncertainties in a realistic collision
between two finite-size spheres (the scaling law strictly applies to
a point mass impacting a planar surface). Otherwise, we need to consider
alternatives. These include (i) a smaller impactor with a higher speed
than is typical of the Kuiper belt, (ii) a more efficient way of
producing debris with smaller bodies, as in a surface-skimming impact, or
(iii) a larger impactor but at an early epoch when such bodies were
more common, with $\fkb\gg 1$. We discuss these cases in turn.

An impactor with radius $\ri = 50$~km and an impact speed of $\vi =
4$--5~km/s would produce enough debris to make the satellites, in
accordance with the scaling law in Equation~(\ref{eq:Miest}).
However, from the impact speed distributions in \citet{delloro2013},
only about 10\%\ of the potential impactors sustain these high
speeds. Thus, the likelihood of Charon encountering a fast-moving
smaller body of this size is about the same as the chance of a giant
impact with a slower-moving large projectile.

The second possibility is a surface-skimming impact involving a small
object ($\ri \lesssim 50$~km) with a highly efficient delivery of mass
to the satellite zone. With a smaller mass, we expect less debris. Yet
an oblique impact can produce a well-directed spray pattern
\citep[e.g.,][]{canup2018}. The low probability of getting just the
right impact geometry may be balanced by the high abundance of smaller
objects. For example, there are five times as many objects with radii
greater than 30~km as those with $\ri > 50$~km (for $q = 4$).

Perhaps the most promising avenue is a scenario where the Charon
impact happens at an early time, not long after the circularization of
the Pluto-Charon binary. Because the Sun's outer protoplanetary disk
was more compact and/or dense during the first $\sim$100~Myr of its
evolution \citep[e.g.,][]{kenyon2002, levison2008}, the primordial
Kuiper belt had many more TNOs than at present. Collisional cascades
among TNOs, driven by gravitational stirring \citep{kl1999b, kb2004b}
or stellar flybys \citep{kb2002a}, along with dynamical
ejections \citep{morbi2008} and the clearing of small
particles by solar radiation and wind, removed more than 99\%\ of the
original mass over a period of a billion years \citep[see
  also][]{booth2009}. Thus, while collisions with large TNOs are rare
today, they were common at early times \citep[see Fig,~1 and
  references therein]{kb2014pc}.

If the early Kuiper belt had a mass that was a hundred times greater
than at present, the factor $\fkb$ in Equation~(\ref{eq:pcollide})
would be similarly enhanced. Following that equation, and assuming
that the timescale for mass loss in the Kuiper belt is 0.5~Gyr, the
likelihood of a collision between Charon and a 100~km TNO is about
30\%. It is also $\sim$1,000 times more likely than the collision that
produced the Pluto-Charon binary itself \citep{kb2014pc}.  Indeed, the
likelihood goes up for other scenarios as well, including the
high-speed, lower-mass impactor, and a small, well-aimed
surface-skimming TNO.

Whatever the scenario for an impact, Charon is less likely a target
than Pluto.  From the geometric cross sections alone, an impact on
Charon is about 25\%\ as likely as one on its big
companion. Gravitational focusing factors does not change this
estimate much, since incoming projectiles are typically moving several
times faster than the escape speed of either binary partner. Despite
that Pluto is the bigger target, the mass of an impactor would need to
be at least an order of magnitude higher to create a reservoir of
material in the satellite zone (Fig.~\ref{fig:aiec}). Given the steep
fall-off of the size distribution of TNOs, an impact between Pluto and
such a large object is less likely than a collision between Charon and
a lower-mass projectile.

\subsection{Comparison with other models}

The Charon impactor scenario has one clear advantage over other pictures 
in which the satellites form from debris produced at that same time as
the Pluto-Charon binary. When satellites form during the main Pluto-Charon
impact, tidal expansion of the binary pushes destructive resonances through 
the satellite zone \citep[e.g.,][]{lith2008b, cheng2014b, smullen2017, woo2018}. 
These resonances probably result in the ejection of at least some of the 
four satellites. Although this problem can be mitigated if satellites or 
their precursors are protected from resonant excitation by collisional damping 
within an circumbinary particle disk \citep{walsh2015, bk2015pc}, this protection
must remain in place until the binary has expanded to its present-day orbit. 
Satellite formation tends to be fast compared with binary expansion \citep[see 
discussion in][]{bk2015pc}. Thus, this model requires some tuning between the 
timing of satellite formation and the binary's orbital expansion. 

An impact with another TNO well after the binary expansion finishes
\citep[which occurs within the first 1~Myr of the binary's
  formation;][]{cheng2014a} avoids these difficulties.
\citet{pires2012} identify this advantage in proposing a collision
between two captured TNOs as a mechanism for producing
the satellites. They conclude that this scenario is unlikely as a
result of the short time that these interlopers remain bound to
Pluto-Charon. \citet{lith2008b} explored a similar pathway using a
swarm of captured planetesimals in the early solar system.

Since we rely on an impact between the binary and a wayward TNO
\citep[see also][]{petit2004, parker2012, nesvorny2019}, our model
also overcomes the problem of the satellite system's survival during
Pluto-Charon's tidal evolution. As long as the impact occurs after
tidal expansion is complete ($\sim$ 1~Myr), but before collisional and
dynamical processes have depleted the solar system of TNOs
\citep[$\sim$ 100~Myr][]{kl1999b, kenyon2002, levison2008, booth2009,
  schlicht2011, schlicht2013, kb2014pc}, there are sufficient
projectiles for an impact and small satellites can grow in an
environment where sweeping resonances are not an issue.

Finally, a variant of the scenario presented here involves multiple
impacts from more common TNOs with radii of $O(10)$~km or less. The
challenge then is for debris particles from each impact to be small
and plentiful enough to settle into a circumbinary disk. If low-mass
impactors contribute to the growth of the satellites or their
precursors in this way, we might expect to see evidence of episodic
accretion of small particles onto Nix and Hydra, the largest of the
satellites. \newhorizons\ imagery does not provide a compelling case
for this scenario, as compared with \cassini\ observations of Saturn's
moon, Pan, with its pronounced equatorial ridge. However, the
complicated spins of the Pluto-Charon moons \citep{showalter2019} may
well erase that kind of evidence of accretion.

\subsection{Observational consequences of a Charon-impact origin}

The main advantage of a Charon-impact origin for Styx \textit{et al.}
is that their formation occurs after the destabilizing sweeping
resonances wrought by the tidal expansion of the binary. Instead, the
resonances are at fixed orbital distances. At these locations, disk
material is cleared, leaving a gap through which nothing can migrate.
It is possible that debris particles can accumulate here and
grow through coagulation. Also, by analogy with dynamics in Saturn's
rings \citep[cf][]{crida2010, bk2013}, a satellite may migrate through
a field of debris particles. It will likely stop at the edge of a gap,
as the supply of solids to drive migration is cut off there. We have
simulated exactly this phenomenon in a circumstellar disk with
gaps \citep[see Fig.~3 therein]{bk2011b}.  We caution that the
circumbinary dynamics may be more complicated, involving
satellite-satellite interactions, if circumbinary exoplanets are any
guide \citep{sutherland2019}.

The cratering record on Charon provides evidence for large impacts in
Charon's distant past \citep{singer2019}.  Careful examination of
Charon's surface reveals evidence for craters (impactors) with
diameters as large as 200~km \citep[40~km; see also][]{schenk2018}.
The largest crater, or other large-scale features like the 450~km dark
polar region in Charon's northern hemisphere, may be the fingerprint
of the event that created the satellites.

\section{Conclusion}

In this ``concerto'' on the formation and history of the Pluto-Charon
binary, we assess the plausibility of an impact between Charon and a
large TNO as the origin of the satellite system. Several possibilities 
generate circumstellar disks with enough mass in the satellite zone to
produce the satellites. A direct hit by a 100~km TNO, if treated as a 
giant cratering event, is likely to eject enough mass to build the 
satellites. A direct hit by a smaller body, with $\ri = 50$~km, will also 
work if it efficiently generates ejecta from Charon's surface. A well-aimed
surface-skimming impact by an even smaller, 30~km TNO may also launch
enough debris into the satellite zone.

To form a circumbinary disk with enough material to build the
satellites, the majority of the mass in the ejecta from the Charon-TNO
impact must be in the form of small objects with radii of roughly ten
meters or less.  Then, collisional damping, accelerated by a
collisional cascade, can settle the debris into a long-lived
circumbinary disk. Collisions can also capture more widely distributed
debris particles into the satellite zone.
{With our hybrid
  \nbody-coagulation code \orch, we demonstrate how these
  process operate together to concentrate ejected particles into a
  dynamically cool disk within the satellite zone, truncated
  beyond Hydra's orbit.}

If most of the ejected mass is in sub-millimeter-size particles, then a 
gas cloud with a tenth the mass of the satellites could have entrained 
the debris. This scenario offers the potential of capturing a large fraction 
of the impact ejecta, keeping it in the satellite zone. Gravitational 
settling in the midplane of a circumbinary disk as the gas disperses 
could result in a reservoir of solids for building the satellites.

{Once solid material settles into the dynamically cool disk,
  collisions and gravity drive satellite growth
  \citep{kb2014pc}. Specific outcomes depend on uncertain parameters
  including the surface density profile, particle size distribution,
  and bulk material properties. Whatever the details, growth is
  unavoidable; satellites like Styx, Nix, Kerberos and Hydra are
  inevitable.}

Whether these scenarios are realistic hinges on the likelihood
that a large TNO could hit Charon. If a big ($\sim$100~km) object were
needed to provide enough mass for the satellites, the impact had to
have taken place within the first billion years of the solar system's
history, when such objects were many times more prevalent than they
are today. Otherwise, an impact with a smaller ($\sim$30--50~km) body is
plausible over the age of the solar system, even with present-day TNO
population. Increasing the efficiency of the delivery of mass to the
satellite system raises the probability that a small body is responsible.

Exploring these possibilities further requires detailed calculations
of the impact \citep[e.g.,][]{arakawa2019} and the damping process
\citep[as in][with the \orchestra\ coagulation code]{kb2014pc}.
Depending on the outcomes of these investigations, the physical
properties of the satellites and their orbits may discriminate between
models where the satellites are remnants of debris produced during the
original giant impact that led to the formation of Pluto-Charon or a
somewhat less energetic (but still powerful) collision between a TNO
and Charon.  One possible discriminant is the proximity of each
satellite to a gap-clearing, mean-motion resonance. The pile-up of
solids near resonances or migration of satellites to gap edges
may depend on the origin of debris in the satellite zone and the
timing of its placement there relative to the tidal expansion of the
binary. Numerical calculations of these processes may offer a path to
choosing among plausible alternatives for the origin of Pluto's small
satellites.

\acknowledgments

We are grateful to M. Geller for suggestions on our original
manuscript.
{We also thank an anonymous referee for comments that led
to further improvements.} NASA provided essential support for this
work through Emerging Worlds program grant NNX17AE24G and a generous
allotment of computer time on the NCCS 'discover' cluster.

\bibliography{planets}{}
\bibliographystyle{apj}

\end{document}